\documentclass[prd,preprint,eqsecnum,nofootinbib,amsmath,amssymb,dvips]{revtex4}
\usepackage{amsmath}
\usepackage{graphicx}
\usepackage{bm}
\usepackage{slashed}
\input epsf

\begin {document}

\newcommand{\be}{\begin{equation}}
\newcommand{\ee}{\end{equation}}
\newcommand{\bml}{\begin{mathletters}}
\newcommand{\eml}{\end{mathletters}}
\newcommand{\bes}{\begin{subequations}}
\newcommand{\ees}{\end{subequations}}

\newcommand{\bi}{\begin{itemize}}
\newcommand{\ei}{\end{itemize}}

\def\centerbox#1#2{\centerline{\epsfxsize=#1\textwidth \epsfbox{#2}}}



\title{Shadow Fermions, Messenger Scalars and Leptogenesis}

\author{Huicheng Guo\footnote{Electronic address:hg4c@virginia.edu} and P. Q. Hung\footnote{Electronic address:pqh@virginia.edu}}

\affiliation{Dept. of Physics, University of Virginia, \\
382 McCormick Road, P. O. Box 400714, Charlottesville, Virginia 22904-4714, 
USA}
\date{\today}

\begin{abstract}
A mechanism for leptogenesis at the electroweak scale was investigated in a 
model of dark energy and dark matter proposed by one of us (PQH).
This model involves an asymptotically free gauge group $SU(2)_Z$ and an axion-like 
particle with an $SU(2)_Z$-instanton-induced potential which yields a scenario for 
the dark energy.
Furthermore, the extended particle content of this $SU(2)_Z$ model contains a possible 
candidate for the cold dark matter, namely, the $SU(2)_Z$ ``shadow'' fermion, 
which couples with the Standard Model lepton through a scalar ``messenger field'' 
carrying both $SU(2)_Z$ and electroweak quantum numbers.
Since these shadow fermions are in a real representation of $SU(2)_Z$,
lepton number can be violated in the Yukawa sector and a lepton number asymmetry 
can be generated in the $SU(2)_Z$ particle's decay which is also $CP$-violating and 
``out-of-equilibrium''.
The asymmetry coming from the interference between the tree-level and one-loop amplitudes 
was calculated for both messenger scalar and shadow fermion decays.
It turns out, in order to have a non-vanishing lepton asymmetry and to be consistent with 
the unitarity condition, some shadow fermions have to decay into lighter messenger scalars, 
hence could be a candidate ``progenitor'' for the lepton asymmetry.
\end{abstract}

\maketitle
\thispagestyle {empty}


\section{Introduction}
The origin of the baryon asymmetry, denoted by the ratio of the net baryon number 
density to the entropy density: $\eta_{B}=(n_{B}-n_{\bar{B}})/s
=6.1 \pm 0.3 \times 10^{-10}$, is one of the most puzzling questions in Cosmology. 
A universe which was initially baryon-antibaryon symmetric will leave a baryon number 
of at least eight orders of magnitude smaller than the previous value. 
A set of criteria which must be satisfied by any model of baryogenesis was laid out 
by Sakharov \cite{sakharov} almost forty years ago for the purpose of calculating 
this asymmetry. Grand Unified Theories (GUT) contain the necessary ingredients
for baryogenesis \cite{gut}: the out-of-equilibrium decay of a massive particle 
which violates baryon number as well as CP.
However, there are several issues with this scenario. The most serious one is the 
presence of electroweak (EW) sphaleron processes at temperatures greater than the 
electroweak scale which conserve $B-L$ but violate $B+L$, where $B$ and $L$ are the
baryon and lepton number respectively.
In these scenarios, the lepton number is associated with Standard Model (SM) 
leptons and the baryon number is associated with SM quarks.
Let us recall that, due to the chiral nature of SM, $B$ and $L$ are violated because 
the SM baryonic current, $J_{\mu}^{B}$ and SM leptonic current, $J_{\mu}^{L}$, 
have an anomaly given by
\be
\label{anomaly}
\partial^{\mu} J_{\mu}^{B} = \partial^{\mu} J_{\mu}^{L}=
(\frac{n_f}{32\,\pi^2})\Big( \, -g^{2}\, W_{\mu \nu}^{a} 
\tilde{W}^{a\,\mu \nu}+ g^{'2} B_{\mu \nu} 
\tilde{B}^{\,\mu \nu} \Big) \,,
\ee
where $W_{\mu \nu}^{a}$ and $B_{\mu \nu}$ are the $SU(2)_L$ and $U(1)_Y$ gauge fields 
respectively, and instanton configurations of the gauge field $W_{\mu \nu}^{a}$ could 
have ${\Delta}B={\Delta}L \neq 0$ as some global tunneling effects which could become 
significant at high temperatures. Such $B+L$ violating ``thermal instanton'' effect 
was first computed in \cite{kuzmin} and referred to as {\em sphaleron} process.
It implies that any $B+L$ asymmetry generated by GUT mechanisms would be ``washed out'' 
by the EW sphaleron processes. It was then realized that one might need 
$B-L$ to be violated itself in order to generate any baryon asymmetry.

What might be the possible sources of $B-L$ violation?

A very promising mechanism under the name of {\em leptogenesis} was proposed 
in which an out-of-equilibrium decay of a heavy Majorana neutrino which 
violates $B-L$ is responsible for a lepton number asymmetry (L-asymmetry) 
\cite{fukugita}, \cite{luty}. If this happens at high enough 
temperatures while the EW sphaleron processes are still in equilibrium, 
this lepton asymmetry can be converted into a baryon asymmetry. 

The aforementioned leptogenesis scenarios have spawned a considerable
amount of very interesting works, especially in connection with
constraints on neutrino masses (see e.g. the excellent review
by Buchm\"{u}ller, Peccei, and Yanagida in \cite{luty}). It goes without
saying that much remains to be done along this path. From an
experimental point of view, the question of whether neutrinos
are Majorana or Dirac is far from being settled, with more
experiments being planned to study this issue.
The attractive and popular see-saw mechanism
which gives rise to small neutrino masses, contains Majorana 
neutrinos, with the heavier ones being candidates for the
leptogenesis scenario. (There are scenarios in which heavy
Dirac neutrinos could be responsible for leptogenesis
\cite{lindner}.) In view of these issues, it might
be interesting to investigate alternative scenarios
of leptogenesis. Could there be a mechanism of leptogenesis in which
the $B-L$ violation comes from the decay of some particle
other than the heavy Majorana neutrino? After all, it is
the SM lepton number violation which is at the heart of the
matter, no matter what its source might be. 
Can one test this new scenario in
terms of its particle physics implications?

There is indeed such a particle as described in \cite{hung2}.
It arises in the construction of a model of dark energy and dark matter \cite{hung2}, 
\cite{su2}. We summarize below the essence of that model in order to motivate the model
of leptogenesis presented in this paper.

As in Refs. \cite{hung2}, \cite{su2}, in this model has proposed an axion-like particle, 
$a_Z$, of a spontaneously broken global $U(1)_{A}^{(Z)}$ symmetry whose potential is 
induced by the instantons of a new unbroken gauge group $SU(2)_Z$.
The $SU(2)_Z$ coupling becomes large at a scale $\Lambda_Z \sim 10^{-3}\,eV$ starting 
from an initial value of the order of the SM couplings at some high energy scale $M$ 
which is much larger than the electroweak scale $\Lambda_{EW}$.
The scenario which was proposed in \cite{hung2}, \cite{su2}, is one
in which $a_Z$ gets trapped in a false vacuum of an instanton-induced
potential with a vacuum energy density $\sim (10^{-3}\,eV)^4$. This model
mimics a universe which is dominated by a cosmological
constant and cold dark matter. 
In fact, the analysis from the Supernova Legacy Survey (SNLS) \cite{snls} fits a flat 
$\Lambda\,CDM$ with a constant equation of state $w = -1.023 \pm 0.090(stat) \pm 0.054(sys) $.
From the observations of WMAP \cite{WMAP}, the combination of WMAP and SNLS data yields 
a constraint $w = -0.967 \begin{array}{c}+0.073 \\ - 0.072 \end{array}$. 
Also as noticed in \cite{WMAP}, even without the prior that the universe is flat, the 
combination of WMAP, large scale structure and supernova data gives $w= -1.08 \pm 0.12$.

As discussed in \cite{hung2}, this $SU(2)_Z$ model, 
besides providing a scenario for the dark energy, contains several other 
phenomenological and cosmological consequences, two of which involve a candidate 
for the cold dark matter and a mechanism  for a new scenario of leptogenesis. 
These aforementioned candidates depend on each other in an interesting way.
As proposed in \cite{hung2} and further explored in \cite{hung4},  a possible source 
of the cold dark matter could be the $SU(2)_Z$ shadow fermions.
These shadow fermions, which transform as $(3,1,0)$ under 
$SU(2)_Z \otimes SU(2)_L \otimes U(1)_Y$, would not have any interaction with the SM 
particles (the visible sector, other than the gravitational one) if it were not for 
the presence of a messenger scalar field $\tilde{\varphi}^{(Z)} =(3,2,-1/2)$. 
As discussed in \cite{hung2}, this presence manifests itself in a variety of ways: 
it could help maintain thermal equilibrium between the $SU(2)_Z$ and SM plasmas 
so that the two sectors possess a common temperature until it drops out of thermal 
equilibrium. Its decay into an SM lepton plus an $SU(2)_Z$ fermion is $CP$-violating 
and hence could possibly generate an L-asymmetry.
The purpose of the this paper is to present a detailed description of this 
new mechanism of leptogenesis. A preliminary version has been presented in \cite{hung3}.

We would like to mention that there exist models of
baryogenesis where there is an asymmetry between SM particles
and e.g. particles that are not affected by the electroweak sphalerons
\cite{kuzmin2} or scalar condensates \cite{dodelson}. Our model
is similar in spirit but is entirely different from the
aforementioned interesting models.

The paper is organized as follows.
First, we will give a brief summary of the salient points of the $SU(2)_Z$ model as 
originally proposed in \cite{hung2}. We notice that the symmetries of the model admit of 
an additional Majorana mass term in the original Lagrangian as appeared in \cite{hung2},
and the shadow fermions can be described using Majorana spinors which are particularly 
convenient in the computations of $CP$-violation for leptogenesis.
So, we will reformulate the theory with shadow fermions in terms of Majorana spinors.

With the so-formed Yukawa interactions in this model, SM leptons could be 
produced in the decays of either shadow fermions or messenger scalars depending on the 
mass order of these $SU(2)_Z$ particles, hence both of them could serve as the ``progenitor'' 
particle whose decay will generate the L-asymmetry when the $CP$-violation and 
out-of-equilibrium conditions are also satisfied. 
We first summarize the zero-temperature results of the $CP$-violation from 
the messenger scalar decay and argue that though each decay channel could have 
$CP$-violation, they will sum to zero if all messenger scalars are heavier than the 
shadow fermions and hence no L-asymmetry could be produced. This implies that some 
unstable shadow fermions might be the more plausible progenitor particles for 
leptogenesis. Hence we proceed with the calculations of $CP$-violation from the 
shadow fermion decay. In these calculations, only SM lepton masses are ignored since 
all other particles involved might have comparable masses of the electroweak scale.
We will take a little digression at this point to discuss how this interesting 
result is consistent with the unitarity condition required by the Boltzmann equations 
of leptogenesis and further argue that unitarity remains self-consistent for 
other possible scenarios that will be considered.

Then, we move on to the analysis of leptogenesis and show that in some specific 
$SU(2)_Z$ scenario, sufficient L-asymmetry could be generated, which in turn, 
is subsequently reprocessed into the observed baryon asymmetry through 
the EW sphaleron process.
Instead of solving the Boltzmann equations which actually have their own 
limitations in these $SU(2)_Z$ scenarios, we will present a more qualitative argument 
based on the main criteria for leptogenesis and asymptotic approximations 
of L-asymmetry at freezeout.
First we outline some general concerns and constraints that need to be satisfied 
for a successful leptogenesis in generic $SU(2)_Z$ scenarios and then focus on 
the shadow fermion decay and investigate two specific cases containing 
one or two messenger scalars.
It turns out, though some non-zero L-asymmetry could be generated with only one light 
messenger scalar, that it is far from  sufficient to account for the currently observed 
baryon asymmetry. Hence it is more feasible to have one more ``heavy'' messenger scalar 
whose mass is much larger than the weak scale. Then we will argue its mass has an 
upper bound and is much less than the GUT scale. 
We end with a brief summary of conclusions and some further discussions of the model.

\section{$SU(2)_Z$ Model}
In this section, we summarize the essential elements of the $SU(2)_Z$
model used in \cite{hung2}, restricting ourselves to the
non-supersymmetric case. First we introduce the particle content as 
appeared in \cite{hung2}. In the originally proposed model, the shadow 
fermions are formulated in forms of Dirac spinors, it turns out to be more convenient 
to use Majorana spinors in the discussions of leptogenesis, 
hence we will lay down the theoretical justification for this transformation.

\subsection{Brief Review of $SU(2)_Z$ particle content}
\label{review}

This $SU(2)_Z$ model is basically an extension from the Standard Model (SM) by 
directly multiplying a new gauge group $SU(2)_Z$ to the SM gauge group sector at 
some high energy scale $M \gg \Lambda_{EW}$. The gauge group is described by:
\be
\label{gauge}
G_{SM} \otimes SU(2)_Z 
\ee
Beside the SM fermions, in the model has proposed some $SU(2)_Z$ fermions which 
could serve as a possible candidate for the cold dark matter (thus referred to as 
{\em shadow} fermions).
Fermion fields transform under the above gauge group as
\be
\label{fermion}
\Psi_{(L,R)}^{SM} = (R_{L,R}, 1) \,;\;\; 
\psi_{i,(L,R)}^{Z} = (1,3) \,,
\ee
where $i=1,2,\cdots$ labels the different families of $SU(2)_Z$ fermion triplets, 
and $R_{L,R}$ denotes the representation of the left-handed and right-handed 
SM fermions under $G_{SM}$. Notice that $SU(2)_Z$ is a vector-like gauge group and 
the shadow fermions are chosen to be triplets `` $3$ of $ SU(2)$'' which is a 
$real$ representation\footnote
{The Lie group $SU(2)$ is locally isomorphic to $SO(3)$, the representation 
`` $3$ of $SU(2)$'' is just the fundamental representation of $SO(3)$ which is real.}
in order to ``slow down'' the evolution of the $SU(2)_Z$ coupling.
Also, fermions of each sector are singlets under the other's gauge group, and the 
two sectors can communicate with each other through some messenger scalar fields 
which carry quantum numbers of both sectors:
\be
\label{messenger}
{\tilde{\varphi}}_{a}^{Z} =({\tilde{\varphi}}_{a}^{Z,0},
{\tilde{\varphi}}^{(Z),-}_{a})=
(1,2,Y_{\tilde{\varphi}^Z}=-1,\,3)\,,
\ee
under $SU(3) \otimes SU(2)_L \otimes U(1)_Y \otimes SU(2)_Z$, 
where $a=1,2,\cdots$ labels the different members of the messenger fields,
and where $Q= T_{3L} + Y/2$.
Since we wish $SU(2)_Z$ gauge symmetry to be unbroken, we will assume 
that the potential for ${\tilde{\varphi}}_{a}^{Z}$ is such that 
$\langle {\tilde{\varphi}}_{a}^{Z} \rangle =0$. 
As a consequence, it will {\em not} contribute to the breaking of the 
electroweak gauge group.

One important remark is in order here. Since the shadow fermions are
in a real representation of $SU(2)_Z$, lepton number can be violated
because $\psi_{R}^Z$ and $(\psi_{L}^Z)^{c}$ are in the same representation.
They can however be either Majorana or Dirac. In particular, lepton number
can be violated and leptogenesis is possible also in the Dirac case.

As in \cite{hung2}, the $SU(2)_Z$ sector also contains  a complex scalar field 
$\bm{\phi}^Z$ which is a singlet under both the SM and $SU(2)_Z$ gauge group and is 
responsible for the shadow fermion masses and a scenario of the dark energy.\footnote
{As shown in \cite{hung2}, the ``axion'', which is the imaginary part of  this complex 
scalar, gets trapped in a false vacuum and yields a scenario for the dark energy.}
At some high scale $M \gg \Lambda_{EW}$, the model exhibits a $U(1)_A^{(Z)}$ global 
symmetry: 
\[ \begin{array}{ccc}
\psi_i^{Z} \rightarrow e^{i\alpha\gamma_5}\,\psi_i^{Z}\,, & 
 \psi_{i,L}^{Z} \rightarrow e^{-i\alpha}\,\psi_{i,L}^{Z}\,, & 
 \psi_{i,R}^{Z} \rightarrow e^{i\alpha}\,\psi_{i,R}^{Z}\,, \\
 \bm{\phi}^{Z} \rightarrow e^{-2i\alpha}\, \bm{\phi}^{Z}\,, &
 l_{L}^{m} \rightarrow e^{i\alpha}\, l_{L}^{m}\,, &
 \tilde{\varphi}_{a}^{Z} \rightarrow \tilde{\varphi}_{a}^{Z}\;.
\end{array} \]
This $U(1)_A^{(Z)}$ symmetry plays an important role in the emergence of an $SU(2)_Z$ 
instanton-induced axion potential which yields a false vacuum and drive the present 
accelerating universe which is assumed to be trapped in this false vacuum. Also, 
what is more relevant here is that the spontaneous breakdown of $U(1)_A^{(Z)}$ gives 
masses to the shadow fermions as $\bm{\phi}^{Z}$ acquires a real
vacuum-expectation-value (VEV):
\be 
\langle\bm{\phi}^Z\rangle\;=\;v_Z \;.
\ee
This VEV is  unconstrained by present particle physics data, 
although a recent model of ``low scale'' inflationary scenario did put 
a constraint on it \cite{inflation}. 
As a result, the masses of $\psi_{i}^{Z}$ are arbitrary. 
Furthermore, in order for $SU(2)_Z$ to be asymptotically free and $SU(2)_Z$ 
coupling $\alpha_Z$ to grow strong (of order unity) at around 
$\Lambda_Z=3 \times10^{-3}\,eV$, so that some stable shadow fermions can be 
confined and form a WIMP cold dark matter, the evolution of $\alpha_Z$ requires 
the presence of two shadow fermions, namely: $\psi_{1}^{Z}$ and $\psi_{2}^{Z}$. 
Interestingly as we will see, this turns out to be in agreement with the 
requirement for a leptogenesis scenario. Also, as argued in \cite{hung2}, 
the most attractive WIMP scenario in our model is one in which 
$\psi_{1}^{Z}$ and $\psi_{2}^{Z}$ are close in mass to each other,
with $m_{\psi^{Z}_{2}} \sim m_{\psi^{Z}_{1}} \sim \Lambda_{EW}$.
The constraint of the running coupling $\alpha_Z$ also implied that at most one 
messenger scalar could have mass as low as of the weak scale $\Lambda_{EW}$, 
and there might be some other messenger scalars with masses much larger than 
$\Lambda_{EW}$ whose appearance will not alter too much the running 
behavior of $\alpha_Z$. As we should see, actually such heavy messenger scalars 
turn out to be needed for a successful leptogenesis scenario.
It is in this specific $SU(2)_Z$ context that we will concentrate our discussion
of leptogenesis later in section \ref{Leptogenesis_specific}.

\subsection{Reformulation in terms of Majorana spinors}
\label{transformation}
The model was originally proposed with shadow fermions of Dirac-type \cite{hung2}, 
it turns out to be much simpler and more straightforward to 
describe the shadow fermions in terms of Majorana spinors. 
Notice that as the shadow fermions $\psi^Z_{i}$ form a real representation of
$SU(2)_Z$, and $(\psi^Z_{i,L})^c$ transforms like a right-handed spinor, one
can write the $SU(2)_L \otimes U(1)_Y \otimes SU(2)_Z$ and global $U(1)_A^{(Z)}$ 
invariant Yukawa interactions as follows

\be
\label{yuk} 
{\cal L}_{yuk}=\sum_{i,a,m} \bm{g}_{a,m}^{(i)}\,
                              \tilde{\varphi}_{a}^Z\,
                              \bar{l}_L^{m}\,(\psi_{i,R}^Z\,+\,(\psi_{i,L}^Z)^{c})\,+\,
               \sum_i\Big(\,k_i\,(\bar{\psi}_{i,L}^Z\,\psi_{i,R}^Z)\,\bm{\phi}^Z\,+\,
                           h_i\,(\,(\psi_{i,R}^Z)^T\,C\,\psi_{i,R}^Z)\,\bm{\phi}^Z\,
                     \Big)\,+\, H.c. \;,
\ee
where $m$ labels the SM lepton families, and where $C=i\,\gamma_2\,\gamma_1$ 
is the charge conjugation Dirac matrix.
The first term\footnote
{In general, the Yukawa couplings for $\psi_{i,R}^Z$ and $(\psi_{i,L}^Z)^{c}$ interacting 
with the SM leptons through $\tilde{\varphi}_{a}^Z$ may be different. Here we assume them 
to be equal for later convenience since there is no obvious reason to insist such a 
difference between the shadow fermions and their anti-partners.} 
in (\ref{yuk}) is the relevant part for leptogenesis, 
and the second term\footnote
{For simplicity, here we neglect the flavour mixing among the shadow fermions.}, 
similar to the see-saw mechanism presented in \cite{seesaw}, 
is responsible for the masses of shadow fermions:
the term containing $k_i$ gives rise to a Dirac-type mass and that containing 
$h_i$ is the Majorana mass term. It is more convenient and neater to express $\psi^Z_i$ 
in terms of two-component Weyl spinors $\chi_i$, $\eta_i$ $\in SL(2,C)$
that are commonly adopted in the formalism of supersymmetry \cite{susy},
\be
\psi^Z_i = \left( \begin{array}{c}
              (\chi_i)_\alpha \\
              (\bar{\eta_i})^{\dot{\alpha}} \end{array} \right) \;,
\ee
as $\bm{\phi}^Z$ acquires a VEV:
$ \langle\bm{\phi}^Z\rangle=v_Z$, the $U(1)_A^{(Z)}$ symmetry was spontaneously
broken, and the relevant parts from the second term in (\ref{yuk}) that give 
masses to the shadow fermions can be expressed in terms of $\chi$, $\eta$ as
\be
\label{Dirac mass} 
{\cal L}_{mass}= -\frac{1}{2}\,\sum_i
                \left( \begin{array}{cc}
                  \chi_i & \eta_i \end{array} \right)
                \left( \begin{array}{cc}
                   0 & |k_i|v_Z \\
                   |k_i|v_Z & 2|h_i|v_Z \end{array} \right)
                \left( \begin{array}{c}
                   \chi_i \\
                   \eta_i \end{array} \right)+ H.c. \;. 
\ee   
The mass matrix is real\footnote
{Though here the Yukawa couplings are in general complex, we could rephase 
the two-component spinors $\chi_i$ and $\eta_i$ to make the mass matrix real.}
 and symmetric, thus can be diagonalized,
\be 
\label{Majoranamass1}
{\cal L}_{mass}= -\frac{1}{2}\,\sum_i
                \left( \begin{array}{cc}
                  \zeta_i & \xi_i \end{array} \right)
                \left( \begin{array}{cc}
                   m_i & 0 \\
                   0 & n_i \end{array} \right)
                \left( \begin{array}{c}
                   \zeta_i \\
                   \xi_i \end{array} \right)+ H.c. \;,
\ee
where $m_i$, $n_i$ are the mass eigenvalues given by\footnote
{The mass eigenvalues are real but $n_i$ could be negative, 
this minus sign has no physical meaning but is just a convention of changing 
the sign of the mass term in the Dirac equation, and this will imply an interchange 
between the definitions of the $u$-spinor and $v$-spinor. However, in practical 
calculations, one can bypass this subtlety by just using signed values of the masses 
as given by (\ref{eigen-mass}).}
\be
\label{eigen-mass} 
\begin{array}{c}m_i\\
                n_i \end{array}
              =|h_i|v_Z\pm\sqrt{|k_i|^2v_Z^2+|h_i|^2v_Z^2}\;.
\ee
It is convenient to define the angle $\theta_i$ by
\be
\label{angle} 
\tan{2\theta_i}\equiv\frac{|k_i|}{|h_i|} \;,
\ee   
in terms of $\theta_i$, the two-component spinor eigenstates $\zeta_i$, $\xi_i$ are
\be
\label{eigen_spinor}
\left( \begin{array}{c}\zeta_i \\ \xi_i \end{array} \right)
=\left( \begin{array}{cc}
                  \sin\theta_i & \cos\theta_i\\
                  \cos\theta_i & -\sin\theta_i \end{array} \right)
            \left( \begin{array}{c}
                   \chi_i \\
                   \eta_i \end{array} \right) \; . 
\ee

Then the new mass eigenstates in Dirac space can be described by the four-component 
Majorana spinors $M_i$ and $N_i$, which are constructed from the above two-component 
spinors $\zeta_i$ and $\xi_i$ respectively, i.e.
\be
\label{def Majorana}
\begin{array}{cc}
M_i = \left( \begin{array}{c}
              \zeta_i \\
              \bar{\zeta_i}\end{array} \right)\,; &\;\;\;
N_i = \left( \begin{array}{c}
        \xi_i \\
        \bar{\xi_i}\end{array} \right) \end{array}\,.
\ee
In forms of these Majorana spinors, the mass term (\ref{Majoranamass1}) 
can be expressed as
\be 
\label{Majoranamass2}
{\cal L}_{mass}=-\frac{1}{2}\,\sum_i
                  \big(\,m_i\,\overline{M}_iM_i\,+\,
                       n_i\,\overline{N}_iN_i \,\big) \,.
\ee
And, the sum of the Dirac spinors of shadow fermions in the first term of (\ref{yuk}) 
can be expressed as
\be
\label{DtoM}
\psi_{i,R}^Z+(\psi_{i,L}^Z)^c
=(\cos\theta_i+\sin\theta_i)M_{i,R}+(\cos\theta_i-\sin\theta_i)N_{i,R}\;.
\ee
By substituting (\ref{DtoM}) into (\ref{yuk}), the relevant part of Yukawa 
interactions responsible for leptogenesis becomes
\be
\label{yuk_new}
{\cal L}_{yuk}= \sum_{i,a,m}\Big(\,g_{a,m}^{(i)}\, 
                            \tilde{\varphi}_{a}^Z\,\bar{l}_L^{m}\,M_{i,R}\,+\,
                              g_{a,m}^{\prime (i)}\,
                            \tilde{\varphi}_{a}^Z\,\bar{l}_L^{m}\,N_{i,R}\,\Big)\;+\;
                             H.c.\;+\;\cdots
\ee
where the new Yukawa couplings:
\begin{eqnarray}
\label{newcoupling1}
 g_{a,m}^{(i)}=(\cos\theta_i+\sin\theta_i)\,\bm{g}_{a,m}^{(i)} \;,\\
\label{newcoupling2}
 g_{a,m}^{\prime (i)}=(\cos\theta_i-\sin\theta_i)\,\bm{g}_{a,m}^{(i)} \;.
\end{eqnarray}        

To complete the transformation, we also need to put the kinetic term of 
the shadow fermions in terms of these Majorana spinors so that 
this transformation could yield a physical equation of motion which is 
crucial for the quantization of the fields \cite{majorana}.
This can be done up to a total derivative, i.e.
\be
\label{kinetic_new}
\bar{\psi}^Z_i\slashed{D}\psi^Z_i \;\longrightarrow\;
   \frac{1}{2}\,\big(\,\overline{M}_i\slashed{D}M_i\,+\,
                       \overline{N}_i\slashed{D}N_i\,\big) \;,
\ee
where the covariant derivative:
$D_\mu=\partial_\mu-ig_Z\bm{T}\cdotp\bm{A}^Z_\mu$, and $ \bm{T}$ is the
generator of the $SU(2)_Z$ gauge group, here $(T_i)_{jk}=i\epsilon_{ijk}$ 
for the $real$ representation `` $3$ of $ SU(2)$''. Actually the decomposition in 
(\ref{kinetic_new}) is neither trivial nor a mere coincidence, it depends crucially
on the form of generators of a real representation. 

From the above transformation, we can see, each Dirac spinor field 
$\psi^Z_i$ can be decomposed into two Majorana spinor fields $M_i$ and 
$N_i$ with the total degrees of freedom unchanged.
In general, the Yukawa couplings in the theory are arbitrary complex numbers.
When $h_i\neq0$ and shadow fermions acquired masses, the mass 
eigenstates will be of Majorana type. 
An interesting case is when the Majorana mass terms are absent as originally 
proposed in \cite{hung2}, i.e. $h_i=0$, 
and the mass term in (\ref{yuk}) could just give the shadow fermion a Dirac 
mass $m_{\psi_i^Z}=|k_i|v_Z$ when $U(1)_A^{(Z)}$ was spontaneously broken.
However from the above transformation, the theory is equally good in terms of 
Majorana spinors. This implies when $h_i=0$ in (\ref{yuk}), the shadow fermions 
can be described using both Dirac and Majorana spinors with equal masses, 
i.e. $m_{\psi_i^Z}=|m_i|=|n_i|=|k_i|v_Z$.
In this case, the transformations of the shadow fermions from Dirac spinors into 
Majorana spinors do not affect the physical consequence of the theory but do 
make the descriptions and calculations simpler.\footnote
{For example, in representing the shadow fermions in a Feynman diagram, 
we will use a doubled-line to represent the Majorana spinor, such a ``doubled'' line 
can be interpreted as two single lines, both representing Dirac spinors with 
opposite particle number.}

To avoid unnecessary complications and be consistent with \cite{hung2}, 
from here on, we will assume 
\be
h_i\,=\,0\;,
\ee 
and for convenience, we will use Majorana spinors to describe the shadow fermions
through out the paper.
Then (\ref{angle}) implies $\theta_i=\pi/4$, 
and (\ref{newcoupling1}), (\ref{newcoupling2}) become
\be
g_{a,m}^{(i)}=\sqrt{2}\,\bm{g}_{a,m}^{(i)}\,;\;\;\;\;\; g_{a,m}^{\prime (i)}=0 \;.
\ee
so, the Majorana spinor $N_i$ completely decoupled\footnote
{This fact relies on the assumption made previously in the first term in (\ref{yuk}) that 
$\psi_{i,R}^Z$ and $(\psi_{i,L}^Z)^{c}$ have equal Yukawa couplings to the SM leptons.}
in the new Yukawa interactions and only $M_i$ will be involved, i.e.
\be
\label{yuk_new2}
{\cal L}_{yuk}= \sum_{i,a,m}\,g_{a,m}^{(i)}\, 
                            \tilde{\varphi}_{a}^Z\,\bar{l}_L^{m}\,M_{i,R}\,+\,
                             H.c.\;+\;\cdots \;.
\ee
For clarity, from now on we will denote the mass of $M_i$ as $m_{M_i}$, i.e.
\be
m_{M_i} \equiv  m_i=|k_i|v_Z=m_{\psi_i^Z} \;,
\ee
and denote the mass of messenger scalar $\tilde{\varphi}_{a}^Z$ as 
$m_{\tilde{\varphi}_{a}^Z}$ whose origin was discussed in \cite{hung2}.
The following analysis of leptogenesis will be based on this simplified Yukawa 
interaction (\ref{yuk_new2}), and all analysis can be carried over to the
general case if necessary, although we expect the conclusion reached in this paper 
to remain the same.

As we can see, the Yukawa interactions in (\ref{yuk_new2}) obviously violate SM 
lepton numbers, the decays of either shadow fermions or messenger scalars 
into SM leptons could generate an L-asymmetry if such decay processes are also 
$CP$-violating and out-of-equilibrium.
We will next present the general computation results and implications of 
the $CP$-violations from the $SU(2)_Z$ particle decays.

\section{$CP-$Violation from $SU(2)_Z$ Particle Decay}
In the content of leptogenesis (or baryogenesis), $CP$ violation is ``measured'' by 
the difference in the rates of a decay mode from its $CP$-conjugate mode. In practice, 
a non-vanishing $CP$-violation could come from the interference between 
the tree-level and one-loop contributions to the decay widths. 
We will first present the results with arbitrary numbers of shadow fermions and 
messenger scalars for later convenience. 

\subsection{Messenger Scalar Decay}
\label{Messenger Scalar Decay}

\begin{figure}
\centerbox{1}{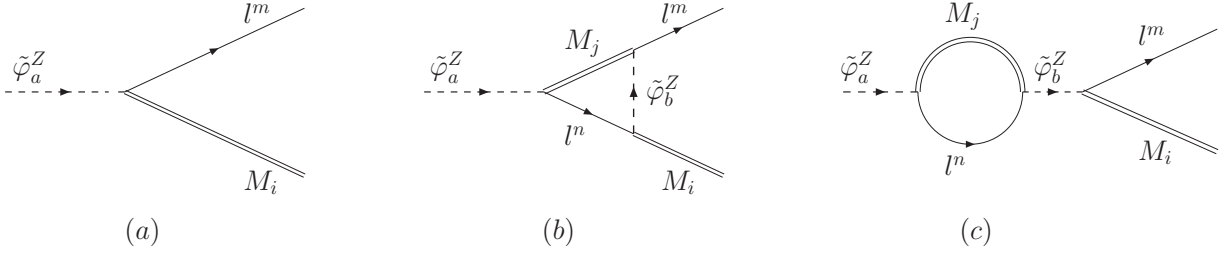}
\caption{\label{fig1}
Messenger scalar $\tilde{\varphi}_{a}^Z$ decay at
(a)tree-level, (b)vertex-one-loop and (c)self-energy-one-loop.
For $\tilde{\varphi}_{a}^{Z,*} \rightarrow M_i+ \bar{l}^m$:
the corresponding diagrams are the same as above except that all the arrows
are reversed.
}
\end{figure} 

First let us consider the messenger scalar $\tilde{\varphi}^Z_a$ decay.
To avoid confusions, some notational conventions are in place.
Throughout this paper, we use $m,n$ to label the different members of SM lepton families, 
and $i,j$ for shadow fermions, and $a,b$ for the messenger scalars as shown in Fig.1,
and we will put a ``tilde'' sign on top of the notations in the case that 
the asymmetry is generated by the decay of $\tilde{\varphi}_{a}^Z$.

The difference in the partial decay rate of $\tilde{\varphi}_a^Z$ is
\be
\label{sdecaydiff}
\Delta\tilde\Gamma_{im}^{a}\equiv
         \Gamma(\tilde{\varphi}_{a}^{Z} \rightarrow M_i+l^m)
        -\Gamma(\tilde{\varphi}_{a}^{Z,*} \rightarrow M_i+ \bar{l}^m) \;,
\ee
a non-vanishing value of $\Delta\Gamma_{im}^{a}$ could come from 
the interference between the tree-level and one-loop contributions 
to the decay widths (as shown in Fig.1), in the leading order,
\be
\label{sdecaydiffint}
\Delta\tilde\Gamma_{im}^{a}=
         \Gamma_{int}(\tilde{\varphi}_{a}^{Z} \rightarrow M_i+l^m)
        -\Gamma_{int}(\tilde{\varphi}_{a}^{Z,*} \rightarrow M_i+ \bar{l}^m) \;.
\ee
It can be splitted into two parts in which the one-loop contribution comes
from the vertex correction (Fig.1(b))and self-energy (Fig.1(c)) respectively,
\be
\label{sdecaysplit}
\Delta\tilde\Gamma_{im}^{a}=\sum_{b,j,n}
                    \Big(\,(\Delta\tilde\Gamma_V^{ab})_{ij \atop mn}\,+\,
                     (\Delta\tilde\Gamma_S^{ab})_{ij \atop mn}\, \Big) \;,
\ee
where, as shown in Fig.1, the indices $a,i,m$ label the ``external'' messenger scalars, 
shadow fermions and SM leptons respectively (that appeared as external lines), 
while $b,j,n$ label the ``internal'' particles (that appeared as internal propagators),
and $(\Delta\Gamma_{V,S}^{ab})_{ij \atop mn}$ can be put in the form
\be
\label{deltagamma}
(\Delta\tilde\Gamma_{V,S}^{ab})_{ij \atop mn}=
    \frac{m_{\tilde{\varphi}_a^Z}}{16\pi}\,
     Im\{(\tilde{G}_{V,S}^{ab})_{ij \atop mn}\}\,
     (\tilde{I}_{V,S}^{ab})_{ij} \;,
\ee
where $(\tilde{G}_{V,S}^{ab})_{ij \atop mn}$ is some product of the Yukawa couplings  
and $(\tilde{I}_{V,S}^{ab})_{ij}$ is the imaginary part of some loop integral times 
the kinematic factor. 
Note that $(\tilde{I}_{V,S}^{ab})_{ij}$ does not carry indices $m,n$ because all
SM leptons are massless.

Eventually, only the total decay difference will account for the L-asymmetry, hence 
we need to sum over all possible final states, that is, sum over $i,\,m$ in 
(\ref{sdecaysplit}) to get the total decay difference
\begin{eqnarray}
\label{sdecaytot}
\Delta\tilde\Gamma_{tot}^{a}&=& \sum_{i,m}\Delta\tilde\Gamma_{im}^{a} \nonumber \\
                            &=& \sum_b\sum_{i,j}\sum_{m,n}
                          \Big( \,(\Delta\tilde\Gamma_V^{ab})_{ij \atop mn}\,+\,
                  (\Delta\tilde\Gamma_S^{ab})_{ij \atop mn}\,\Big) \;. \nonumber \\
\end{eqnarray}
For convenience, we define
\be
\label{gsum}
(\tilde{G}_{V,S}^{ab})_{ij}\equiv\sum_{m,n}(\tilde{G}_{V,S}^{ab})_{ij \atop mn}\;,
\ee
then we can substitute (\ref{deltagamma}) into (\ref{sdecaytot}) and use (\ref{gsum})
to get the more compact form
\be
\label{deltagammatot1}
\Delta\tilde\Gamma_{tot}^{a}= \frac{m_{\tilde{\varphi}_a^Z}}{16\pi}\sum_b\sum_{i,j}
                \Big(\,Im\{(\tilde{G}_{V}^{ab})_{ij}\}(\tilde{I}_{V}^{ab})_{ij}\,+\,
                 Im\{(\tilde{G}_{S}^{ab})_{ij}\}(\tilde{I}_{S}^{ab})_{ij}\,\Big) \;.
\ee 

Now we need the detailed expressions for the factors $\tilde{G}_{V,S}^{ab}$ 
and $\tilde{I}_{V,S}^{ab}$.
$(\tilde{G}_{V,S}^{ab})_{ij \atop mn}$ can be easily read off from the diagrams, and 
it gives directly
\begin{eqnarray}
\label{sgvform}
(\tilde{G}_{V}^{ab})_{ij}= \sum_m(g_{a,m}^{(i),*} g_{b,m}^{(j)}) 
                   \sum_n(g_{b,n}^{(i),*} g_{a,n}^{(j)}) \;,\\
\label{sgsform}
(\tilde{G}_{S}^{ab})_{ij}= \sum_m(g_{a,m}^{(i),*} g_{b,m}^{(i)}) 
                   \sum_n(g_{b,n}^{(j),*} g_{a,n}^{(j)}) \;.
\end{eqnarray} 
Some useful properties of $\tilde{G}_{V,S}^{ab}$ are
\begin{eqnarray}
\label{sgvproperty}
(\tilde{G}_{V}^{ab})_{ij}=(\tilde{G}_{V}^{ab})_{ji}^*=(\tilde{G}_{V}^{ba})_{ij}\;,\\
\label{sgsproperty}
(\tilde{G}_{S}^{ab})_{ij}=(\tilde{G}_{S}^{ab})_{ji}^*=(\tilde{G}_{S}^{ba})_{ij}^*\;.
\end{eqnarray}
The first equalities in (\ref{sgvproperty}) and (\ref{sgsproperty}) imply that 
$Im\{(\tilde{G}_{V,S}^{ab})_{ij}\}=0$ when $i=j$, thus $\Delta\tilde\Gamma_{tot}^{a}=0$,
and we immediately reached an important fact that no asymmetry can be generated if 
there is only one shadow fermion triplet. 
Amusingly, this is in agreement with \cite{hung2}, \cite{su2} where two of such triplets
are needed to ``slow down'' the evolution of the $SU(2)_Z$ gauge coupling from a scale 
of $\mathcal{O}(\Lambda_{GUT})$ to the electroweak scale. Also notice that 
$(\tilde{G}_{V,S}^{ab})_{ij}$ is $U(3)$ invariant in the lepton family space, 
so these $CP$-violations cannot be eliminated by rephasing or mixing the SM leptons.

Before write down the expression for $(\tilde{I}_{V,S}^{ab})_{ij}$, it is convenient 
to define some mass squared ratios (as in (\ref{sxy}) in the Appendix):
\[ \begin{array}{ccc}
   \tilde{s}_b^a=\big(\frac{m_{\tilde{\varphi}_b^Z}}{m_{\tilde{\varphi}_a^Z}}\big)^2\,,&
   \tilde{d}_i^a=\big(\frac{m_{M_i}}{m_{\tilde{\varphi}_a^Z}}\big)^2\,,&
   \tilde{d}_j^a=\big(\frac{m_{M_j}}{m_{\tilde{\varphi}_a^Z}}\big)^2\,. \end{array}
\]
In terms of these mass squared ratios, the vertex-one-loop contribution can be 
expressed as the product of some kinematic factor and the $V$-function,
\begin{eqnarray}
\label{Svintegral}
(\tilde{I}_{V}^{ab})_{ij} 
&=& -\frac{1}{4\pi} \sqrt{\tilde{d}_i^a\tilde{d}_j^a}\,
    V(\tilde{s}_b^a,\tilde{d}_i^a,\tilde{d}_j^a) \nonumber \\
&=& -\frac{1}{4\pi} \sqrt{\tilde{d}_i^a\tilde{d}_j^a}\,\Big\{\, 
    V_1(\tilde{s}_b^a,\tilde{d}_i^a,\tilde{d}_j^a)\,\theta(1-\tilde{d}_j^a)\,+\,
    V_2(\tilde{s}_b^a,\tilde{d}_i^a,\tilde{d}_j^a)\,
    \theta\Big(1-\frac{\tilde{s}_b^a}{\tilde{d}_i^a}\Big)
                \,\Big\}\,\theta(1-\tilde{d}_i^a) \;,\nonumber \\
\end{eqnarray}
where, as summarized in the Appendix, the function $V$ comes from the imaginary part of 
the vertex-one-loop integral, and it could be expressed as a sum of the functions
$V_1$ and $V_2$ which are given by (\ref{V1A}), (\ref{V2A}) in the Appendix, and 
$\theta(\cdotp)$ is the step function as usual.
The self-energy-one-loop contribution is easier to get,
\be
\label{Ssintegral}
(\tilde{I}_{S}^{ab})_{ij}
= -\frac{(1-\tilde{d}_i^a)^2(1-\tilde{d}_j^a)^2}{4\pi(1-\tilde{s}_b^a)}\,
   \theta(1-\tilde{d}_i^a)\,\theta(1-\tilde{d}_j^a) \;.
\ee
With $\tilde{G}_{V,S}^{ab}$ and $\tilde{I}_{V,S}^{ab}$ given above, we obtained 
an exact expression for the total decay difference $\Delta\tilde\Gamma_{tot}^{a}$ 
in (\ref{deltagammatot1}).

From the above results, we can demonstrate that the shadow fermions cannot 
all be stable, i.e. some of them have to decay into some lighter messenger scalars. 
If the shadow fermions were all stable, the mass order would be
\[ m_{\tilde{\varphi}_{a,b}^Z}>m_{M_{i,j}} \;,\]
i.e. $\tilde{d}_{i,j}^a<1$ and ${\tilde{s}_b^a}/\tilde{d}_{i,j}^a>1$,
so that it is energetically forbidden for the shadow fermions to decay into messenger 
scalars. Hence the $V_2$ term in (\ref{Svintegral}) vanished due to the step function, 
and only $V_1$ contributes $(\tilde{I}_{V}^{ab})_{ij}$. As could be seen from (\ref{V1A}) 
in the Appendix, $V_1(s,x,y)$ is invariant under the interchange of variables 
$x=\tilde{d}_i^a$ and $y=\tilde{d}_j^a$. 
Also from (\ref{Ssintegral}) we can see $(\tilde{I}_{S}^{ab})_{ij}$ is also invariant 
under the interchange of $\tilde{d}_i^a$ and $\tilde{d}_j^a$. 
These correspond to the physical processes with the interchange between the external and 
internal shadow fermions, i.e. $i \leftrightarrow j$. Hence we have
\[  \begin{array}{cc}
    (\tilde{I}_{V}^{ab})_{ij}=(\tilde{I}_{V}^{ab})_{ji }\,,\;\; & \;\;
    (\tilde{I}_{S}^{ab})_{ij}=(\tilde{I}_{S}^{ab})_{ji} \;, \end{array}
\]
but from (\ref{sgvproperty}) and (\ref{sgsproperty}),
\[ \begin{array}{cc}
    Im\{(\tilde{G}_{V}^{ab})_{ij}\}=-Im\{(\tilde{G}_{V}^{ab})_{ji}\}\,,\;\; &\;\;
    Im\{(\tilde{G}_{S}^{ab})_{ij}\}=-Im\{(\tilde{G}_{S}^{ab})_{ji}\}\,, \end{array} 
\]
thus, $\Delta\tilde\Gamma_{tot}^{a}$ vanishes after summing over $i,j$, and no
asymmetry will be generated. This indicates eventually at least one 
shadow fermion should have a mass heavier than some messenger scalars so that 
it may decay into a lighter messenger scalar and an SM lepton. As we will see later, 
this result is consistent with the unitarity condition required by the setup of the 
Boltzmann equations describing the dynamics of leptogenesis.


\subsection{Shadow Fermion Decay}

\begin{figure}
\centerbox{1}{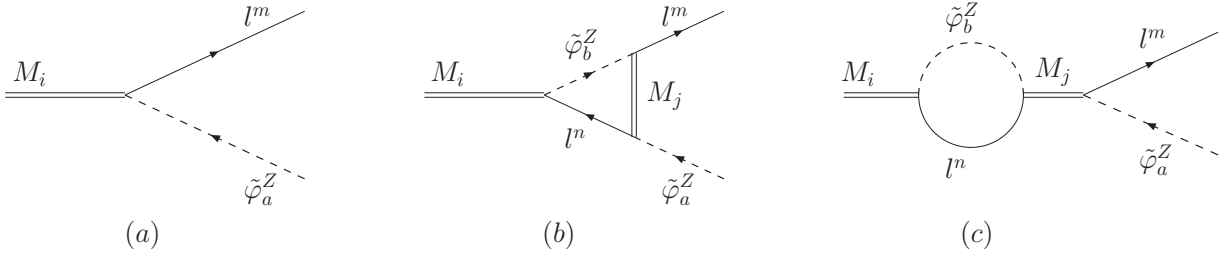}
\caption{\label{fig2}
Shadow fermion $M_i$ decay at
(a)tree-level, (b)vertex-one-loop and (c)self-energy-one-loop.
}
\end{figure} 

Similar to the analysis in the messenger scalar decay, the difference in the partial 
decay rates of the shadow fermions described by the Majorana spinor  $M_i$ is
\be
\label{fdecaydiff}
\Delta\Gamma_{am}^{i}\equiv
         \Gamma(M_i \rightarrow \tilde{\varphi}_{a}^{Z,*}+l^m)
        -\Gamma(M_i \rightarrow \tilde{\varphi}_{a}^{Z}+\bar{l}^m)\;,
\ee
and a non-vanishing value of $\Delta\Gamma_{am}^{i}$ could come from 
the interference between the tree-level and one-loop contributions 
to the decay widths (as shown in Fig.2). Similar to the previous section,
the total decay rate difference can be written as
\be
\label{deltagammatot2}
\Delta\Gamma_{tot}^{i}= \frac{m_{M_i}}{32\pi}\sum_j\sum_{a,b}
                      \Big(\,Im\{(G_{V}^{ij})_{ab}\}(I_{V}^{ij})_{ab}\,+\,
                       Im\{(G_{S}^{ij})_{ab}\}(I_{V}^{ij})_{ab}\,\Big)\;,
\ee
where
\begin{eqnarray}
\label{fgvform}
(G_{V}^{ij})_{ab}= \sum_m(g_{a,m}^{(i),*} g_{b,m}^{(j)}) 
                   \sum_n(g_{b,n}^{(i),*} g_{a,n}^{(j)}) \;,\\
\label{fgsform}
(G_{S}^{ij})_{ab}= \sum_m(g_{a,m}^{(i),*} g_{a,m}^{(j)}) 
                   \sum_n(g_{b,n}^{(j),*} g_{b,n}^{(i)}) \;,
\end{eqnarray} 
and they satisfy
\begin{eqnarray}
\label{fgvproperty}
(G_{V}^{ij})_{ab}=(G_{V}^{ji})_{ab}^*=(G_{V}^{ij})_{ba} \;,\\
\label{fgsproperty}
(G_{S}^{ij})_{ab}=(G_{S}^{ji})_{ab}^*=(G_{S}^{ij})_{ba}^* \;.
\end{eqnarray}
We still have $Im\{(G_{V,S}^{ij})_{ab}\}=0$ when $i=j$, thus 
$\Delta\Gamma_{tot}^{i}=0$, this leads to the same conclusion that L-asymmetry 
cannot be generated if there's only one shadow fermion triplet in agreement 
with \cite{hung2}, \cite{su2}. 
And still, $(G_{V,S}^{ij})_{ab}$ is $U(3)$ invariant in the lepton family space.

Now, it is convenient to define mass squared ratios as
\[ \begin{array}{ccc}
   s_j^i=\big(\frac{m_{M_j}}{m_{M_i}}\big)^2,&
   d_a^i=\big(\frac{m_{\tilde{\varphi}_a^Z}}{m_{M_i}}\big)^2,&
   d_b^i=\big(\frac{m_{\tilde{\varphi}_b^Z}}{m_{M_i}}\big)^2, \end{array}
\]
then, the expressions for $(I_{V,S}^{ij})_{ab}$ are almost the same as 
(\ref{Svintegral}) and (\ref{Ssintegral}) up to some different kinematic factors and 
an interchange between the indices $i \leftrightarrow a$, $j \leftrightarrow b$, i.e.
\begin{eqnarray}
\label{Fvintegral}
(I_{V}^{ij})_{ab}
&=& \frac{1}{4\pi}\sqrt{s_j^i}\,V(s_j^i,d_a^i,d_b^i) \nonumber \\
&=& \frac{1}{4\pi}\sqrt{s_j^i}\,\Big\{\, 
    V_1(s_j^i,d_a^i,d_b^i)\,\theta(1-d_b^i)\,+\,
    V_2(s_j^i,d_a^i,d_b^i)\,\theta\Big(1-\frac{s_j^i}{d_a^i}\Big)
                           \,\Big\}\,\theta(1-d_a^i) \;,\nonumber \\
\end{eqnarray}
and
\be
\label{Fsintegral} 
(I_{S}^{ij})_{ab}=\frac{(1-d_a^i)^2(1-d_b^i)^2}{8\pi(1-s_j^i)}\,
                  \theta(1-d_a^i)\,\theta(1-d_b^i) \;.
\ee
where $V_1$, $V_2$ still have the same functional form as given by 
(\ref{V1A}), (\ref{V2A}) in the Appendix.
Comparing the results of messenger scalar decay with shadow fermion decay, 
we can see, the identical functional forms come from the crossing symmetry while the 
different kinematic factors are due to the different initial and final states in question.

Before applying these results to analyze the leptogenesis in this $SU(2)_Z$ model, 
let us take a glance at the unitarity condition.


\section{Unitarity}
\label{unitarity}

\begin{figure}
\centerbox{1}{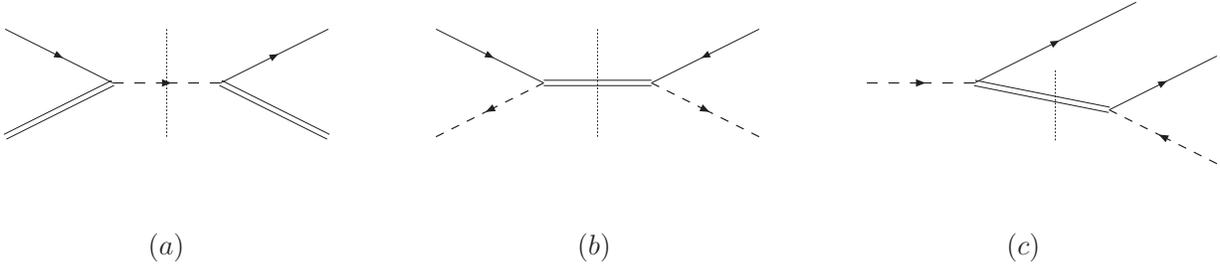}
\caption{\label{fig3}
``Cuts''at real-intermediate-states
}
\end{figure} 

This section is somewhat a digression from the main subject of the paper, 
thus we will make it concise.
But it is worth discussing since it provides a different point of view of 
one important conclusion proved in \ref{Messenger Scalar Decay} stating that the 
shadow fermions cannot all be stable. Furthermore, we will argue that unitarity 
remains as a self-consistent condition for the possible leptogenesis scenarios 
that will be considered in detail later.

Unitarity condition is a general constraint in the setup of Boltzmann equations 
which give a more quantitative and dynamical description for leptogenesis 
(and the earlier baryogenesis).
As argued in \cite{wolfram} (by the so-called Boltzmann's $H$-theorem), no asymmetry 
between the particle and anti-particle can exist when a system is in equilibrium 
provided the theory is CPT-invariant and unitary. 
In the standard formalism of Boltzmann equations for leptogenesis or baryogenesis, 
the unitarity condition was maintained by subtracting out the contribution from the 
{\em real intermediate state} (RIS) of $s-$channel scattering amplitude to the second 
order of the relevant couplings, which has been already counted 
in the tree-level inverse decay and decay processes at the same order.
The RIS contributions can be represented by a ``cut'' on the $s-$channel 
propagators as shown in Fig.3.

Now we can revisit the conclusion that the shadow fermions cannot be all 
stable in the spirit of the unitarity condition. If all shadow fermions were
stable, the asymmetry can only be generated by the decay of messenger scalars
and the RIS has to be produced by the  ``cut'' on the $s-$channel propagators
of the messenger scalars as shown in Fig.3(a). However, this $s-$channel
scattering is a lepton-number-conserving process, thus could not enter the simplified
Boltzmann equation of leptogenesis which includes only the lepton-number-violating 
processes. Putting this in simple words, if $\Delta\Gamma_{tot}^{a}\neq 0$,
L-asymmetry could be generated even in equilibrium and the unitarity (to the second 
order of the relevant couplings) is violated. Hence, in order to generate L-asymmetry,
at least one shadow fermion has to be unstable and at least one is stable to 
account for the dark matter candidate. 
This suggests the mass order (in the minimal particle content):
\be
\label{massorder1}
m_{M_2}\;>\;m_{\tilde{\varphi}_1^Z}\;>\;m_{M_1} \;.
\ee
How is unitarity maintained in this case?
First notice that still no asymmetry will be generated from $\tilde{\varphi}_1^Z$ decay 
into SM lepton and lighter shadow fermion $M_1$.
This is because, first from (\ref{fgvproperty}) and (\ref{fgsproperty}), 
$Im\{(\tilde{G}_{V,S}^{ab})_{ij}\}=0$ 
if $i=j=1$, thus the heavier shadow fermion $M_2$ has to enter the loop; but then, 
both $V_1$ and $V_2$ in (\ref{Svintegral}) vanish because the step functions vanish
due to the mass order (\ref{massorder1}).
So in this particle content, L-asymmetry could only be generated from the 
shadow fermion $M_2$ decay, and the unitarity condition can be enforced by subtracting  
out the RIS from the $s-$channel scattering process via $M_2-$exchange which could be 
L-violating as shown in Fig.3(b). 

(\ref{massorder1}) provides the minimal particle content with all the desired features. 
However, as will be argued later, this minimal content fails to generate sufficient 
L-asymmetry that is needed to account for the currently observed baryon asymmetry.
This problem could be resolved by adding one more heavy messenger scalar, say 
$\tilde{\varphi}_2^Z$, to the theory. As mentioned earlier, it needs to be ``heavy'', 
with its mass much larger than the electroweak scale, in order to satisfy the constraint 
for the running $SU(2)_Z$ coupling. The mass order is
\be
\label{massorder2}
m_{\tilde{\varphi}_2^Z} \,\gg\, m_{M_2}\,>\,m_{\tilde{\varphi}_1^Z}\,>\,m_{M_1} \;.
\ee
In this content, the lepton-number asymmetry could be generated not only in the decay 
of $M_2$ but also in the heavy messenger scalar $\tilde{\varphi}_2^Z$ decay, since this 
time $V_2$ (in (\ref{Svintegral}) for $\tilde{\varphi}_2^Z$ decay) will not vanish 
identically when interchange the shadow fermions $M_1$ and $M_2$. 
However, the lighter messenger scalar $\tilde{\varphi}_1^Z$ decay is excluded 
for the similar reason as mentioned above.
This seems problematic since there is no ``L-violating RIS'' for  $\tilde{\varphi}_2^Z$ 
decay, however, it turns out that the unitarity can still be restored by subtracting 
out the RIS as shown in Fig.3(c), actually it takes care of $M_2$ decay and 
$\tilde{\varphi}_2^Z$ decay simultaneously.
These suggest that unitarity is a self-consistent condition of the theory and can indeed 
be maintained dynamically in various cases as it should be.


\section{General Concerns of $SU(2)_Z$ Leptogenesis}
Following the standard analysis of leptogenesis which is essentially the same as the idea 
of baryogenesis \cite{kolb}, it is useful to define the asymmetry factor characterizing 
the efficiency of L-asymmetry production as
\be
\epsilon\equiv\frac{\Delta\Gamma_{tot}}{\Gamma_{tot}} \;,
\ee
where $\Delta\Gamma_{tot}$ is given by (\ref{deltagammatot1}) or (\ref{deltagammatot2})
for $\tilde{\varphi}_a^Z$ or $M_i$ decay, and $\Gamma_{tot}$ is the total decay rate. 
To the leading order, it is sufficient to evaluate  $\Gamma_{tot}$ 
for $M_i$ or $\tilde{\varphi}_a^Z$ decay at tree-level,
\be
\label{fdecayrate}
\Gamma_{tot}^{i}=\frac{m_{M_i}}{32\pi}\,
                   \sum_{a,m}{|g_{a,m}^{(i)}|}^2\,(1-d_a^i)^2\,
                   \theta(1-d_a^i) \;,
\ee
\be
\label{sdecayrate}
\tilde\Gamma_{tot}^{a}=\frac{m_{\tilde{\varphi}_a^Z}}{16\pi}\,
                   \sum_{i,m}{|g_{a,m}^{(i)}|}^2\,(1-\tilde{d}_i^a)^2\,
                   \theta(1-\tilde{d}_i^a) \;,
\ee
thus, for $M_i$ or $\tilde{\varphi}_a^Z$ decay:
\be
\label{fasym}
\epsilon^{(M_i)}=\frac{\sum_j\sum_{a,b}
                    (Im\{(G_{V}^{ij})_{ab}\}\,(I_{V}^{ij})_{ab}+
                     Im\{(G_{S}^{ij})_{ab}\}\,(I_{S}^{ij})_{ab})}
                  {\sum_{a,m}{|g_{a,m}^{(i)}|}^2\,(1-d_a^i)^2\,
                   \theta(1-d_a^i)} \;;
\ee
\be
\label{sasym}
\epsilon^{(\tilde{\varphi}_a^Z)}=\frac{\sum_b\sum_{i,j}
                    (Im\{(\tilde{G}_{V}^{ab})_{ij}\}\,(\tilde{I}_{V}^{ab})_{ij}+
                     Im\{(\tilde{G}_{S}^{ab})_{ij}\}\,(\tilde{I}_{S}^{ab})_{ij})}
                  {\sum_{i,m}{|g_{a,m}^{(i)}|}^2\,(1-\tilde{d}_i^a)^2\,
                   \theta(1-\tilde{d}_i^a)} \;.
\ee
A non-vanishing value of the asymmetry factor $\epsilon$ reflects two 
essential ingredients for leptogenesis: lepton number violation and $CP$-violation. 
The third necessary ingredient is the ``out-of-equilibrium-decay''. The primary
condition for a departure from thermal equilibrium is the requirement that 
the particle's decay rate  $\Gamma_{D}$ is {\em less than} the expansion rate 
$H = 1.66\, g_{*}^{1/2} T^2/m_{pl}$, where $g_{*} \simeq 114$, is the effective
number of degrees of freedom (including $SU(2)_Z$ light degrees of freedom). 
Here, $\Gamma_D$ is given by (\ref{fdecayrate}) or (\ref{sdecayrate}). 
If the masses of $SU(2)_Z$ particles are not very close such that $d_a^i$ and 
$\tilde{d}_i^a$ are small and could be ignored, it is convenient to define
\be
\alpha_a^{(i)}=\frac{1}{4\pi}\sum_{m}{|g_{a,m}^{(i)}|}^2 \;,
\ee
and $\Gamma_D$ can be approximated as 
\be
\Gamma_{D} \;\sim\; \alpha_{D}\, m_{D} \;,
\ee
where $\alpha_{D} \sim \alpha_a^{(i)}$, and $m_D=m_{M_i}$ or $m_{\tilde{\varphi}_a^Z}$, 
is the mass of the decaying particle.

In order to make the whole picture more transparent, we will first outline 
some basic concerns and asymptotic behaviors of the asymmetry at freezeout as in the 
generic GUT scenarios, and then focus on more specific features of this $SU(2)_Z$ 
leptogenesis scenario at the electroweak scale.
As with \cite{hung3}, \cite{kolb}, we introduce the quantity $K$ defined as the ratio 
of the particle's decay rate to the expansion rate at temperature $T \sim m_D$:
\be
\label{K}
K \equiv \frac{\Gamma_{D}}{2\,H}\Big|_{T=m_D}
\sim \;\frac{\alpha_{D}\,m_{pl}}{3.3\,g_{*}^{1/2}\,m_D}
\sim \;\big(\frac{\alpha_D}{m_D}\big)\times 10^{17} \;,
\ee
where in the last expression $m_D$ is in $GeV$.
The maximal asymmetry in a scenario is obtained when $K \ll 1$, the progenitor 
particles are overabundant and depart from thermal equilibrium; and if the 
rates of all damping processes are also much less than the expansion rate 
and thus can be ignored (referred to as the ``far-out-of-equilibrium'' regime).  
Then, the maximal net lepton-number to entropy ratio, or more precisely, the net 
$B-L$ density per comoving volume at freezeout is
\be
\label{maxasymm}
\eta_{B-L}^{max.}\; \sim \; -\frac{\epsilon}{g_{*}} \;.
\ee
If the currently observed baryon asymmetry was mostly generated in this scenario 
through the sphaleron process, i.e. $\eta_B \sim 0.5\,\eta_{B-L}$ \cite{kuzmin} 
(or more precisely: $\eta_B \sim 0.35\,\eta_{B-L}$ \cite{harvey}), 
with $g_{*} \simeq 114$, a rough constraint on $\epsilon$ is found to be
\be
\label{epslower}
-\epsilon \;> 10^{-7} \;.
\ee 
When $K \ll 1$, the progenitor particle is long-lived and its decay is out of 
equilibrium. Since the time when they decay is $t \sim \Gamma^{-1}_D$ and 
since $T \propto 1/\sqrt{t}$, the temperature at the time of decay is found to be 
(using (\ref{K})) $T_D \sim K^{1/2}\,m_D$ \cite{kolb}. 
For this scenario to be effective, i.e. a conversion of an L-asymmetry 
into a baryon asymmetry through the electroweak sphaleron process, 
one has to make sure that the decay occurs at a temperature greater than 
the critical temperature $T^* \sim 100\,GeV$ above which the sphaleron processes 
are in thermal equilibrium.
From this, it follows that $K$ cannot be arbitrarily small and
has a lower bound coming from the requirement $T_D > T^*$ \cite{hung3}. 
One obtains the rough bounds for $K$:
\be
\label{lower}
1>\,K\, > \;\big(\frac{100\,GeV}{m_D}\big)^2 \,.
\ee
If leptogenesis took place at the weak scale, we can take an upper bound for 
the mass of the progenitor particle to be $10^3\,GeV$, this would give: $0.01<K<1$.
Some remarks are in order concerning the upper bound for $K$ in (\ref{lower}), 
where unity is not an exact upper bound. In other words, some L-asymmetry may still 
be generated even when the progenitor's decay was not ``far-out-of-equilibrium''.
However, as $K$ increase, the damping processes become more and more important, 
when $K>1$ but not too different from unity and the inverse decay is the dominant 
damping process (referred to as the weak washout regime), the net lepton number 
asymmetry will be diluted approximately by a factor of $1/K$ and  has to be 
compensated by an increase in the asymmetry factor $\epsilon^{(i,a)}$. That is,
in the weak washout regime, (\ref{maxasymm}) should be modified by
\be
\label{maxasymm2}
\eta_{B-L}\; \sim \; -\frac{\epsilon}{g_{*}\,K} \;.
\ee
When $K \gg 1$ and $2\leftrightarrow2$ scattering is the dominant damping process
(referred to as the strong washout regime), the asymmetry at freezeout will be 
exponentially suppressed and eventually washed out. 

These asymptotic behaviors are generic in the usual GUT scenarios 
\cite{kolb} and more rigorous solutions could be found by solving the coupled 
Boltzmann equations which describe these out-of-equilibrium processes.
However in this $SU(2)_Z$ model, due to the ``low energy'' scale and the ``big gap'' 
in the orders of Yukawa couplings as we will see, it turns out to be more subtle here 
than in the generic GUT scenarios which is at much higher scale and whose couplings 
are of similar orders.
In order for leptogenesis to occur at such ``low energy'' as of the weak scale, if we 
use the rough bound $m_D < 10^3\,GeV$, from (\ref{K}), $K \sim 1$ implies 
\be
\alpha_D \;<\; 10^{-14} \;,
\ee
the ``decay-couplings'' $\alpha_D$ needs to be very small in order for the decay 
processes to be ``out-of-equilibrium''; on the other hand, some couplings that 
enter the interference terms needs to be relatively very large in order to generate 
sufficient asymmetry. The asymmetry factor $\epsilon$ given by (\ref{fasym}) or 
(\ref{sasym}) can be characterized in the form
\be
\label{epsform}
\epsilon=4\pi\alpha_A(\sin{\kappa_V}\,f_V+\sin{\kappa_S}\,f_S) \;,
\ee
where $\alpha_A$ and $\kappa_{V,S}$ are the characteristic Yukawa coupling and the 
$CP$-violating phase that dominate the asymmetry factor. 
And $f_{V,S}$ are numerical factors coming 
from the loop integrals which turn out to be bounded above for reasonable 
masses of $SU(2)_Z$ particles. If we take $f_{V,S}<1$ for the moment, 
then (\ref{epslower}) and (\ref{epsform}) implies 
\be
\alpha_A \;>\; 10^{-8} \;,
\ee
and $\sin{\kappa_{V,S}}$ should be negative. Hence 
\be
\label{biggap}
\alpha_A \;\gg\; \alpha_D \;.
\ee
More detailed form for $\alpha_A$ and $\alpha_D$ will be given shortly. Note that, 
though (\ref{biggap}) was derived in the ``far-out-of-equilibrium'' regime, 
it still holds in the weak washout regime, since both $\alpha_D$ and $\alpha_A$ will 
need to be modified by multiplying the same factor of $K$, i.e. 
$\alpha_D<10^{-14}\,K$ and $\alpha_A>10^{-8}\,K$.

Because of this special feature of Yukawa couplings, more care need to taken in this 
$SU(2)_Z$ leptogenesis scenario, but the basic strategy is similar. 
First, one needs to compare the damping rates with the expansion rate. 
A damping process will be in equilibrium if its rate is much larger than the 
expansion rate and hence will wash out all L-asymmetry.
If all damping rates are not too large compared to the expansion rate, we could further
determine the asymptotic approximations of the L-asymmetry at freezeout by comparing the 
inverse decay rate with the $2\leftrightarrow2$ scattering processes. 
Also, to address the validity of one underlining assumption in the simplified 
Boltzmann equation that the $SU(2)_Z$ plasma and SM plasma have a common temperature, 
one needs to compare the expansion rate with the lepton-number-conserving 
scattering processes that equilibrize the temperature between the two sectors. 
When the rates are comparable, simplified Boltzmann equation will become unreliable. 

Guided by these general concerns, now we can apply the obtained results to  
investigate some specific scenarios of the $SU(2)_Z$ model and extract 
more information about the Yukawa couplings and the masses of $SU(2)_Z$ particles 
from the constraints mentioned earlier.

\section{Leptogenesis in the specific $SU(2)_Z$ model}
\label{Leptogenesis_specific}
 
In this paper, our main interest is to present a leptogenesis scenario near the 
electroweak scale which is also linked to the concern of dark energy and dark matter.
As already argued, at least one shadow fermion has to be unstable and its decay into 
messenger scalar and SM lepton will give rise to the L-asymmetry.
So from here on, we will mainly focus on the decay of shadow fermion $M_2$ and 
consider about having two shadow fermions and one or two messenger scalars which are the 
most feasible scenarios to accommodate with the dark energy and dark matter \cite{hung2}.
In either case, the main decay channels for leptogenesis are
\[ M_2 \;\longrightarrow \;\tilde{\varphi}_{1}^{Z,*}\,+\,l \;,\]
and the $CP$-conjugate modes. It follows immediately
\be
\label{alphaD}
\alpha_D \;\sim \;\alpha_1^{(2)} \;.
\ee

From the antisymmetry property: $Im\{(G_{V,S}^{ij})_{ab}\}=0$ when $i=j$ and by setting 
$i=2$, $j=1$, $a=1$ in (\ref{fasym}), the asymmetry factor from $M_2$-decay becomes
\be
\label{fasym2}
\epsilon^{(M_2)}=\frac{\sum_{b}
                    (Im\{(G_{V}^{21})_{1b}\}\,(I_{V}^{21})_{1b}+ 
                     Im\{(G_{S}^{21})_{1b}\}\,(I_{S}^{21})_{1b})}
                  {\sum_{m}{|g_{1,m}^{(2)}|}^2\,(1-d_1^2)^2} \;,
\ee
where, as before, $d_1^2=m^2_{\tilde{\varphi}_1^Z}/m^2_{M_2}$ and the index $b$ labels 
the messenger scalars in the loop. 

In what follows, we first examine the case containing only one light messenger scalar with 
the mass order given by (\ref{massorder1}), in which we will find, sufficient L-asymmetry 
cannot be generated. Then we investigate the extended version by adding one more heavy  
messenger scalar with the mass order given by (\ref{massorder2}).
Though the decay of the heavy  messenger scalar could also have a non-vanishing 
asymmetry factor, as we will see, such decays will be in equilibrium and hence will not 
produce L-asymmetry.


\subsection{One Messenger Scalar}
First, consider the simplest case with only one messenger scalar $\tilde{\varphi}_1^{Z}$, 
thus $b=1$ only. From the previous discussions in \ref{review}, \cite{hung2} and \ref{unitarity},
the masses of the ``light'' $SU(2)_Z$ particles are of similar orders of $\Lambda_{EW}$ 
and have to follow the mass order given by (\ref{massorder1}),
hence in this case, we have the mass condition
\be
\label{masscondition1}
\mathcal{O}(10^3\,GeV)\;>\;m_{M_2}\,>\,m_{\tilde{\varphi}_1^Z}\,>\,m_{M_1} \,
\sim \,\mathcal{O}(10^2\,GeV)\;.
\ee
Also, from (5.6), $Im\{(G_{S}^{21})_{11}\}=0$ for $a=b=1$, thus the self-energy contribution 
in the second term in (\ref{fasym2}) vanishes, and $\epsilon^{(M_2)}$ becomes
\be
\label{fasym3}
\epsilon^{(M_2)}=
     \Big(\frac{Im\{(G_{V}^{21})_{11}\}}{\sum_{m}{|g_{1,m}^{(2)}|}^2}\Big)\,
     \Big(\frac{(I_{V}^{21})_{11}}{(1-d_1^2)^2}\Big) \;.
\ee
To make the parameter-dependence more transparent, define the characteristic quantities
as in (\ref{epsform}), substitute (\ref{fgvform}) in (\ref{fasym3}) 
and set $i=2$, $j=1$, $a=b=1$, we have
\be
\label{alpA}
4\pi\,\alpha_A\sin{\kappa_V} =
 \frac{Im\{(G_{V}^{21})_{11}\}}{\sum_m {{|g_{1,m}^{(2)}|}^2}}
=\frac{Im\{\sum_m{g_{1,m}^{(2),*} g_{1,m}^{(1)}}\sum_n{g_{1,n}^{(2),*} g_{1,n}^{(1)}}\}}
      {\sum_m {{|g_{1,m}^{(2)}|}^2}}\;,
\ee
and
\be
\label{fV}
f_V=\frac{(I_{V}^{21})_{11}}{(1-d_1^2)^2}\;.
\ee
Then $\epsilon^{(M_2)}$, which should satisfy the general constraint (\ref{epslower}), 
can be characterized in the form
\be
\label{epsform1}
-\epsilon^{(M_2)}=4\pi\alpha_A(-\sin{\kappa_V})\,f_V>10^{-7}\;.
\ee
From (\ref{alpA}) we can see, the magnitude of $|g_{1,m}^{(2)}|$ cancels in numerator and 
denominator, hence $\alpha_A$ is of the same order as $\alpha_1^{(1)}$,
\be
\alpha_A \sim \alpha_1^{(1)}=\frac{1}{4\pi}\sum_{m}{|g_{1,m}^{(1)}|^2}
\ee
Furthermore, the mass order (\ref{massorder1}) implies 
$d_1^2=m^2_{\tilde{\varphi}_1^Z}/m^2_{M_2}<1$ and 
$s_1^2/d_1^2=m^2_{M_1}/m^2_{\tilde{\varphi}_1^Z}<1$, thus both $V_1$ and $V_2$ will 
contribute to $f_V$, i.e. 
\be
\label{fV2}
f_V =\frac{\sqrt{s_1^2}\,\{ V_1(s_1^2,d_1^2,d_1^2)+V_2(s_1^2,d_1^2,d_1^2)\}}
          {4\pi(1-d_1^2)^2}\;,
\ee
where $(V_1+V_2)$ takes the form as given by (\ref{V3A}) in the Appendix. 
$f_V$ was found to be non-negative and bounded above: 
\be
0\;\leq \;f_V\;<\;0.05
\ee
The above bounds of $f_V$ hold for arbitrary mass values as long as the mass order 
(\ref{massorder1}) is satisfied.
Then from the constraint (\ref{epsform1}) and take $-\sin{\kappa_V}\sim 1$,
we get
\be
\alpha_A \;\sim \;\alpha_1^{(1)}\;>\;10^{-7}\;,
\ee
recall $\alpha_D \sim \alpha_1^{(2)} \sim 10^{-14}$, we have,
\be
\label{alphagap1}
\alpha_1^{(1)} \;\gg \;\alpha_1^{(2)}\;.
\ee
There seems to be one unnatural  ``hierarchy'' problem with (\ref{alphagap1}) because 
it says that such widely separated Yukawa couplings are carried by the shadow fermions 
with rather similar masses $m_{M_1} \sim m_{M_2}$.
Actually, even if we bear this uncomfortable fact, L-asymmetry will be washed out 
completely by the damping process of the inverse-decay: 
$l+M_1 \rightarrow \tilde{\varphi}_1^Z $. 

At the temperature $T \sim m_{M_2} \sim m_{\tilde{\varphi}_1^Z}$,
\be 
\Gamma_{ID}^{(\tilde{\varphi}_1^Z)} \;\simeq \;\Gamma_{D}^{(\tilde{\varphi}_1^Z)} 
\;\sim \;\alpha_1^{(1)}\,m_{\tilde{\varphi}_1^Z} \;.
\ee
We can compare it with the expansion rate following (\ref{K}),
\be
\label{inverseK}
\frac{\Gamma_{ID}^{(\tilde{\varphi}_1^Z)}}{2\,H}\Big|_{T=m_{\tilde{\varphi}_1^Z}}\;
\sim \;\frac{\alpha_1^{(1)}\,m_{pl}}{3.3\,g_{*}^{1/2}\,m_{\tilde{\varphi}_1^Z}}\;
\sim \;\big(\frac{\alpha_1^{(1)}}{m_{\tilde{\varphi}_1^Z}}\big) \times 10^{17}\;,
\ee
where $m_{\tilde{\varphi}_1^Z}$ is in $GeV$ as in (\ref{K}).
If sufficient asymmetry were to be produced, as argued before, we need 
$\alpha_1^{(1)}>10^{-7}$, and if we take $m_{\tilde{\varphi}_1^Z} < 10^3GeV$ from 
(\ref{masscondition1}), (\ref{inverseK}) implies
\be 
\frac{\Gamma_{ID}^{(\tilde{\varphi}_1^Z)}}{2\,H}\Big|_{T=m_{\tilde{\varphi}_1^Z}}\;
      >\;10^7\;,
\ee
hence the inverse-decay $\;l+M_1 \rightarrow \tilde{\varphi}_1^Z\;$ is in equilibrium 
and the L-asymmetry, if any, will be washed out completely.  

As we already see, the main problem here is that in this minimal particle content, 
the masses of all $SU(2)_Z$ particles are of the order of $\Lambda_{EW}$, so 
all damping processes are energetically allowed at the temperature of this scale. 
On the other hand, a ``big gap'' in the orders of the Yukawa couplings is inevitable 
in order for leptogenesis to take place. ``Small'' couplings are needed for 
``out-of-equilibrium'' decay and ``large'' couplings are needed for sufficient asymmetry. 
Then the damping processes involving the ``large'' couplings ought to be in equilibrium 
and damp away any pre-existing L-asymmetry. Inspired by these facts, the simplest 
solution is to add one more heavy messenger scalar to the theory,\footnote
{Actually, one could also add one more heavy shadow fermion to the theory, 
but this will ruin the running behavior of $SU(2)_Z$ coupling and is not compatible 
with a GUT scenario.} which is also in agreement with \cite{hung3}.

\subsection{Two Messenger Scalars}
With one more heavy messenger scalar $\tilde{\varphi}_2^{Z}$, now the mass order of the 
$SU(2)_Z$ particles is given by (\ref{massorder2}), and still, 
$m_{M_1} \sim m_{M_2} \sim m_{\tilde{\varphi}_1^Z} \sim \Lambda_{EW}$, 
hence in this case, we take the mass condition to be
\be
\label{masscondition2}
m_{\tilde{\varphi}_2^Z}\,\gg\; \mathcal{O}(10^3\,GeV)\;>\,
m_{M_2}\,>\,m_{\tilde{\varphi}_1^Z}\,>\,m_{M_1} \,\sim\, \mathcal{O}(10^2\,GeV)\;.
\ee
To avoid the unnatural ``hierarchy'' problem as appeared in (\ref{alphagap1}),
it is reasonable to assume that $\tilde{\varphi}_2^{Z}$ carries the ``large'' couplings 
and $\tilde{\varphi}_1^{Z}$ carries the ``small'' couplings, i.e.
\be
\label{assumedgap}
\alpha_A \,\sim\, \alpha_2 \;\gg\; \alpha_D \,\sim\, \alpha_1\;,
\ee
where $\alpha_2 \sim \alpha_2^{(1)} \sim \alpha_2^{(2)}$ and  
$\alpha_1 \sim \alpha_1^{(1)} \sim \alpha_1^{(2)}$.
With (\ref{assumedgap}), the contribution from $b=1$ in (\ref{fasym2}) can be ignored, 
because this corresponds the diagram with $\tilde{\varphi}_1^Z$ in the loop and has an 
amplitude $\propto \alpha_1$, thus is much smaller than the one with $\tilde{\varphi}_2^Z$ 
in the loop ($b=2$) which has an amplitude $\propto \alpha_2$.
Still, the self-energy contribution $(I_{S}^{21})_{12}$ in (\ref{fasym2}) vanished 
due to the step function, hence in this case, $\epsilon^{(M_2)}$ becomes
\be
\label{fasym4}
\epsilon^{(M_2)}=
     \Big(\frac{Im\{(G_{V}^{21})_{12}\}}{\sum_{m}{|g_{1,m}^{(2)}|}^2}\Big)\,
     \Big(\frac{(I_{V}^{21})_{12}}{(1-d_1^2)^2}\Big)\;.
\ee
Similarly, we can put $\epsilon^{(M_2)}$ in the characteristic form as in (\ref{epsform1}),
\be
\label{epsform2}
-\epsilon^{(M_2)}=4\pi\alpha_2(-\sin{\kappa_V})\,f_V\;>\;10^{-7}\;,
\ee
where now, by setting $i=2,\,j=1$ and $a=1,\,b=2$, and noticing that $V_1$ vanished 
due to the step function, we have
\be
\label{alpA1}
4\pi\,\alpha_2\sin{\kappa_V} =
 \frac{Im\{(G_{V}^{21})_{12}\}}{\sum_m {{|g_{1,m}^{(2)}|}^2}}
=\frac{Im\{\sum_m{g_{1,m}^{(2),*} g_{2,m}^{(1)}}\sum_n{g_{2,n}^{(2),*} g_{1,n}^{(1)}}\}}
      {\sum_m {{|g_{1,m}^{(2)}|}^2}}\;,
\ee
and
\be
\label{fV1}
f_V=\frac{(I_{V}^{21})_{12}}{(1-d_1^2)^2}
 =\frac{\sqrt{s_1^2}\,\{ V_2(s_1^2,d_1^2,d_2^2)\}}
       {4\pi(1-d_1^2)^2}\;.
\ee
Since 
$m_{\tilde{\varphi}_2^Z}\gg m_{M_2}$, $d_2^2=m^2_{\tilde{\varphi}_2^Z}/m^2_{M_2}\gg 1$, 
$V_2$ can be simplified to the asymptotic form and $f_V$ becomes
\be
f_V\simeq \frac{\sqrt{s_1^2}\,(1-s_1^2/d_1^2)^2}{8\pi\,d_2^2}\;.
\ee
If regard $f_V$ as a function of $s_1^2$, one can find its maximum
\be
f_V^{max}\simeq \frac{0.3\,\sqrt{d_1^2}}{8\pi\,d_2^2}\;,
\ee
when $s_1^2=d_1^2/5$. 
Combine with (\ref{epsform2}), substitute $d_1^2=m^2_{\tilde{\varphi}_1^Z}/m^2_{M_2}$ 
and $d_2^2=m^2_{\tilde{\varphi}_2^Z}/m^2_{M_2}$, we obtain
\be
\label{massconstraint}
\alpha_2(-\sin{\kappa_V})\,
\frac{0.15\,m_{\tilde{\varphi}_1^Z}\,m_{M_2}}{m_{\tilde{\varphi}_2^Z}^2}\;>\;10^{-7}\;.
\ee
If take $\alpha_2 \sim 1$, $-\sin{\kappa_V}\sim 1$, and use 
$m_{\tilde{\varphi}_1^Z}\sim m_{M_2} < 10^3GeV$ from (\ref{masscondition2}), 
we can get an upper bound for the mass of the heavy messenger scalar:
\be
\label{phi2massupper}
m_{\tilde{\varphi}_2^Z} \;<\; 1.5\times 10^6GeV\;.
\ee

Equipped with (\ref{massconstraint}), we can easily see that no L-asymmetry will be 
generated in the decay of heavy messenger scalar $\tilde{\varphi}_2^Z$. Following 
(\ref{K}) (and we will use the mass values in $GeV$ throughout the calculations),
\be
\frac{\Gamma_{D}^{(\tilde{\varphi}_2^Z)}}{2\,H}\Big|_{T=m_{\tilde{\varphi}_2^Z}}\;
\sim \;\big(\frac{\alpha_2}{m_{\tilde{\varphi}_2^Z}}\big) \;\times 10^{17}\;,
\ee
substitute $\alpha_2$ from (\ref{massconstraint}) into above, we get
\be
\frac{\Gamma_{D}^{(\tilde{\varphi}_2^Z)}}{2\,H}\Big|_{T=m_{\tilde{\varphi}_2^Z}}\;>\;
\big(\frac{m_{\tilde{\varphi}_2^Z}}{m_{\tilde{\varphi}_1^Z}\,m_{M_2}}\big)
    \times 10^{11}\;\gg\; 10^8\;,
\ee 
hence $\tilde{\varphi}_2^Z$ decay is in equilibrium and all the L-asymmetry ought to 
be generated from $M_2$ decay at temperatures of the order of the electroweak scale.

Now we will need to examine if sufficient L-asymmetry could be produced in this 
scenario by comparing the damping rates to the expansion rate. 
The thermally-averaged damping rate of a generic $2\leftrightarrow2$ scattering at 
temperature $T$ can be approximated as
\be
\label{Scattrate}
\Gamma_{S} \;\sim\; n \langle \sigma |v|\rangle \; \sim\;
n\,\frac{\alpha\,\alpha^{'}\,T^2}{(T^2+m^2)^2}\;,
\ee 
where $n$ is the number density of the primary particle under concern, $\alpha$ and 
$\alpha^{'}$ are the couplings at the two vertices, and $m$ is the particle's mass 
in the propagator.  

First note that, from the constraint for out-of-equilibrium decay, here we still 
have (\ref{alphaD}): $\alpha_D \sim \alpha_1 \sim 10^{-15}$, so the rates of 
$2\leftrightarrow2$ scattering processes that involve only $\tilde{\varphi}_1^Z$ 
are of the order $\mathcal{O}(\alpha_D^2)$, thus is much smaller than the main decay 
rates and can be ignored. 
Now look at the scattering processes involving $\tilde{\varphi}_2^Z$ which carries 
the large coupling $\alpha_2$. Note that $\tilde{\varphi}_2^Z$ can only appear as 
external particles in L-violating $2\leftrightarrow2$ scattering processes, 
its density at temperature $T \sim m_{M_2}\ll m_{\tilde{\varphi}_2^Z}$ will be highly 
suppressed due to the Boltzmann blocking factor:
\be
n_{\tilde{\varphi}_2^Z} \;\propto \;\exp\{-m_{\tilde{\varphi}_2^Z}/T\}
\ee
So even the scattering cross section is enhanced by a factor of 
$\mathcal{O}(\alpha_2/\alpha_1)$ or $\mathcal{O}((\alpha_2/\alpha_1)^2)$ 
comparing to those involving  $\tilde{\varphi}_1^Z$ , such damping processes are 
even more negligible when $m_{\tilde{\varphi}_2^Z}/m_{M_2} > 100$
at temperature $T \sim m_{M_2}$.
Similarly, the inverse-decays of  $\tilde{\varphi}_2^Z$ will be blocked as well 
at this temperature. 

Hence, at $T \sim m_{M_2}$, the dominant damping processes are the inverse-decays of
$\tilde{\varphi}_1^Z$ and $M_2$, i.e. $\;l+M_1 \rightarrow \tilde{\varphi}_1^Z\;$ and 
$ \;l+\tilde{\varphi}_{1}^{Z,*} \rightarrow M_2 \;$, the rates approximately are
\be
\Gamma_{ID}^{(\tilde{\varphi}_1^Z)} \simeq \Gamma_{D}^{(\tilde{\varphi}_1^Z)} 
\sim \alpha_1\,m_{\tilde{\varphi}_1^Z}\;,
\ee
\be
\Gamma_{ID}^{(M_2)}\simeq \Gamma_{D}^{(M_2)}\sim \alpha_1\,m_{M_2}\;,
\ee
notice also $m_{\tilde{\varphi}_1^Z} \sim m_{M_2}$, we get
\be
\Gamma_{ID}^{(\tilde{\varphi}_1^Z)}\;\sim\;\Gamma_{ID}^{(M_2)} \sim \Gamma_{D}^{(M_2)}\;,
\ee
the damping rates of inverse decay processes are of the same order as the main decay rates. 
So the asymptotic approximation (\ref{maxasymm2}) for the weak washout regime is applicable here 
just as in the generic GUT scenario \cite{kolb}, i.e.
\be
\label{weakwashout}
\eta_{B-L} \;\sim\; \frac{-\epsilon^{(M_2)}}{g_{*}\,K} \;.
\ee 
Then the constraint (\ref{massconstraint}) should be further modified as
\be
\label{massconstraint2}
\alpha_2\,\frac{m_{\tilde{\varphi}_1^Z}\,m_{M_2}}{m_{\tilde{\varphi}_2^Z}^2}
\;>\; \Big\{ \begin{array}{cc}
        10^{-6}\,,&\;10^{-2}<K<1 \\ 10^{-6}\,K\,,&\;K>1 \end{array}
\ee
hence we get one main constraint (\ref{massconstraint2}) for leptogenesis 
in this specific scenario.

In principle, we could further apply Boltzmann equations to solve for the L-asymmetry at 
freezeout in this scenario and get a more precise constraint between the masses of the 
$SU(2)_Z$ particles and Yukawa couplings as in (\ref{massconstraint2}). 
However, besides the heavy messenger scalars $\tilde{\varphi}_2^Z$ which had already 
started decoupling at temperature $T \sim m_{\tilde{\varphi}_2^Z} \gg \Lambda_{EW}$, 
other $SU(2)_Z$ particles $M_1$, $M_2$ and $\tilde{\varphi}_1^Z$ will also start 
decoupling from SM sector at the temperature  $T \sim \Lambda_{EW}$ since 
their masses are of this scale, then the $SU(2)_Z$  plasma will begin to possess its own 
temperature $T_Z \neq T_{SM}$.
Or more precisely, to examine if we could assume a common temperature for the two sectors 
in the simplified Boltzmann equations as described in\cite{wolfram}, we need to look at 
the L-conserving processes, and a common temperature (i.e. $T_Z= T_{SM}$) can be assumed 
when such processes are in equilibrium. 

As argued before, the dominant one among these L-conserving processes is the
$2\leftrightarrow2$ scattering via $\tilde{\varphi}_2^Z-$exchange. This time there is no 
Boltzmann blocking factor since $\tilde{\varphi}_2^Z$ appears only in the propagator. 
From (\ref{Scattrate}), the rate approximately is
\be
\label{phi2Scattrate}
\Gamma_{S}^{(\tilde{\varphi}_2^Z)} \; \sim\;
T^3\,\frac{(\alpha_2)^2\,T^2}{(T^2+m_{\tilde{\varphi}_2^Z}^2)^2}\;,
\ee 
at $T \sim m_{M_2} \ll m_{\tilde{\varphi}_2^Z}$, it can be approximated as
\be 
\Gamma_{S}^{(\tilde{\varphi}_2^Z)}\;\sim\;
\frac{(\alpha_2)^2\,m_{M_2}^5}{m_{\tilde{\varphi}_2^Z}^4}\;,
\ee
substitute $\alpha_2$ from (\ref{massconstraint}) into above and follow (\ref{K}), we get
\be
\frac{\Gamma_{S}^{(\tilde{\varphi}_2^Z)}}{2\,H}\Big|_{T=m_{M_2}} 
\;\sim\; \frac{10^5\,m_{M_2}}{m_{\tilde{\varphi}_1^Z}^2}\;,
\ee
where all masses are in $GeV$.
If use $m_{M_2} \sim m_{\tilde{\varphi}_1^Z} \sim \Lambda_{EW}$, the above ratio is about 
$\mathcal{O}(10^{2}) \sim \mathcal{O}(10^{3})$, and $\Gamma_{S}^{(\tilde{\varphi}_2^Z)}$ 
became rather comparable to the expansion rate already at $T \sim \Lambda_{EW}$. 
This implies though a common temperature between the $SU(2)_Z$ and SM sectors might still 
be achieved, the whole $SU(2)_Z$ sector will start decoupling when the temperature further 
drops and the assumption that $T_Z= T_{SM}$ will become unreliable. To address the problem,
we have to examine the out-of-equilibrium and decoupling behaviors of the other 
$SU(2)_Z$ particles beside the progenitor particle $M_2$ and SM lepton.
Another problem in this scenario is that, around $T \sim \Lambda_{EW}$, the EW sphaleron 
process might become inefficient in converting the lepton asymmetry into a baryon 
asymmetry since its critical temperature is also at the electroweak scale.
Concerning these complications , a separate treatment with  more thorough analysis might be 
appropriate.
 

\section{Conclusions}
In this paper, we presented a scenario of leptogenesis at the electroweak scale in a 
model of the dark energy and dark matter. This amounts to the efforts in investigating 
the possible simple connections between the candidates for leptogenesis and dark matter. 
In this model, new candidates for dark matter was proposed, namely, the shadow fermions 
in a new gauge group sector extended from the Standard Model. These shadow fermions 
communicate with the SM leptons through the so called messenger field, and it is their 
communications that provide a possible mechanism for leptogenesis. 
We extracted some information about the range of couplings and the masses of these 
shadow particles by comparing with the currently observed baryon asymmetry.
First, in order to have leptogenesis and confinement of the light shadow fermions, 
it is crucial for the shadow fermions to form a real representation of the extended 
gauge group, hence these shadow fermions can be described using Majorana spinors even without 
a Majorana mass term.
Also it was found, in order to have a non-vanishing asymmetry, the shadow fermions cannot 
all be stable, i.e. some shadow fermions have to be able to decay into SM leptons and 
light messenger scalars and thus have masses heavier than those product messenger scalars.
Furthermore, this requirement incapacitates the lighter messenger scalars to produce 
L-asymmetry and hence justifies the shadow fermions to be a more suitable progenitor for 
leptogenesis in this scenario. However, the minimal particle content with two shadow fermions 
and one light messenger scalar fails to generate sufficient L-asymmetry, and one more 
heavy messenger scalar with mass much larger than the electroweak scale has to be put into 
the story. Interestingly, if sufficient L-asymmetry were to be produced in this extended 
content, the heavy messenger scalar cannot be too heavy with a mass much less than the 
GUT scale. Hence this leptogenesis scenario is completely at much lower energy scale.
As might be noticed, except for the energy scale, this scenario shares some similarities 
with the popular scenario involving heavy Majorana neutrinos. This casts interesting theoretical 
implications of the important roles that might be played by the Majorana fermions in leptogenesis 
and dark matters.

Last but not least, the favored scenario with two messenger scalars --- one ``light'' 
($m_{\tilde{\varphi}_1^Z} \sim \mathcal{O}(\Lambda_{EW})$) and one ``heavy'' 
($m_{\tilde{\varphi}_2^Z} \sim \mathcal{O}(10^6\,GeV)$) --- has interesting implications at the 
LHC. Since $\tilde{\varphi}_1^Z$ is ``light'' enough, it
can be produced at the LHC and since it carries the ``small coupling'', it will be long-lived and will 
leave interesting signatures in the detector \cite{hung3}.

\begin{acknowledgments}
This work is supported in parts by the US Department of Energy under grant No. 
DE-A505-89ER40518 and the Dissertation Year Fellowship of University of Virginia. 

\[\]

\end{acknowledgments}

\appendix
\begin{center}
    {\bf APPENDIX}
  \end{center}

Here we briefly summarize the computation technique used in evaluating the imaginary 
part of the loop integral. We will mainly focus on the more complicated case of 
vertex-one-loop integral, the computation for the self-energy case will be mentioned 
briefly at the end.

\begin{figure}
\centerbox{0.8}{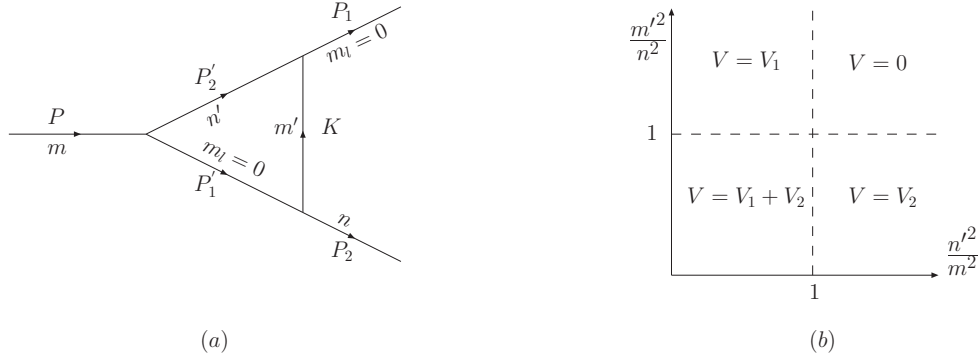}
\caption{\label{fig4}
(a) Vertex-one-loop diagram\;\;  (b) V-function}

\end{figure} 

The main difficulty arose here is that we can only ignore the SM lepton masses.
The central integral to be computed turned out to be the same for both messenger scalar 
decay and shadow fermion decay, hence, as represented in Fig.4(a), we do not 
distinguish the particle type and use only solid line with arrow to indicate the 
momentum flow. 

The main integral from the loop diagram (as shown in Fig.4(a)) to be evaluated 
is the following:
\be
\label{LoopIntegral_Im}
I(m,m';n,n') = 32\,\pi\,\bm{Im}\,\bigg\{\,i
 \int\frac{d^4K}{(2\,\pi)^4}\frac{P_1\cdot P'_1}
     {(K^2-{m'}^2+i\epsilon)\,({P'_2}^2-{n'}^2+i\epsilon)\,({P'_1}^{2}+i\epsilon)} \bigg\}\;,
\ee
where, as labeled in Fig.4(a), $m_l=0$ denotes massless SM leptons; $m,m'$ and $n,n'$ 
are the masses of $SU(2)_Z$ particles and denote the same particle type respectively. 
For example, if  $m,m'$ denote the masses of shadow fermions, then $n,n'$ 
denote the masses of messenger scalars, and vice versa. 
In obtaining (\ref{LoopIntegral_Im}), we used Wick theorem directly to extract out 
the Feynman rules for Majorana spinors which agree with the results proved in 
\cite{majoranarule}.

To evaluate (\ref{LoopIntegral_Im}), we first apply Feynman parameter method to put 
the loop integral in the standard form, then use the formula
\be
\label{ID}
\frac{1}{X-i\,\epsilon}=\bm{Pr}\Big(\frac{1}{X}\Big)+i\pi\,\delta\big(X\big)\;.
\ee
Since we are only interested in the imaginary part, the principal part of the integral 
can be discarded. That is, only the $\delta-$function term will contribute. 
Alternatively, one could also do the integral directly and find out the negative domain 
of the variable in the logarithmic function. In either approach\footnote
{Their equivalence can be checked by integrating by parts.}, $i\epsilon$ description 
plays a crucial role in making the logarithmic function analytic in the complex plane 
with ``branch cut''. Here we picked the easier way by using (\ref{ID}).

To present the general results, first let us define the mass squared ratios as
\be
\label{sxy}
\begin{array}{ccc}
   s=\big(\frac{m'}{m}\big)^2, &
   x=\big(\frac{n}{m}\big)^2, &
   y=\big(\frac{n'}{m}\big)^2. \end{array}
\ee
It turns out to be convenient to define the $V-$function which could be splitted in terms 
of functions $V_1$ and $V_2$ as follows,
\begin{eqnarray}
\label{Vfunction}
V(s,x,y) &\equiv& \big\{I(m,m';n,n')\,(1-x)\big\} \,\theta(1-x) \nonumber \\
&=& \big\{\,V_1(s,x,y)\,\theta(1-y)\,+\,
            V_2(s,x,y)\,\theta\big(1-\frac{s}{x}\big)\big\}\,\theta(1-x) \nonumber \\
\end{eqnarray}
where $V_1$ and $V_2$ are given by:
\be
\label{V1A}
V_1(s,x,y)=-(1-x)(1-y)\,+\,(1+s-x-y)\ln\Big(\frac{1+s-x-y}{s-xy}\Big)\;,
\ee
\be
\label{V2A}
V_2(s,x,y)=(1-x)(1-\frac{s}{x})\,+\,(1+s-x-y)\ln\Big(\frac{\frac{s}{x}-y}{1+s-x-y}\Big)\;,
\ee
and their sum is
\be
\label{V3A}
V_1+V_2=(1-x)(y-\frac{s}{x})\,-\,(1+s-x-y)\ln(x)\;.
\ee
As we can see, the function $V$ contains several step-functions and hence will take various 
forms depending on the different mass orders as illustrated in Fig.4(b), however, it can be 
shown that $V$ is non-negative and continuous despite of the discontinuity form the 
step functions.
These step functions come from the integration of the $\delta-$function as appeared in 
(\ref{ID}) and could be interpreted as if using the ``cutting-rule'' method to evaluate 
the imaginary part of the vertex-one-loop integral, which however, is more involved than 
the method used here.

The calculations for self-energy-one-loop contribution is quite standard and hence will not 
be detailed. The results could be obtained by using both the method mentioned above and 
the ``cutting-rule'' method. And dimension regularization is needed in the method used here 
and the step functions arose from finding the negative domain of the variable in the 
logarithmic function while keeping  $i\epsilon$ description.



\begin{thebibliography}{99}

\bibitem{sakharov} A. D. Sakharov, JETP Lett. {\bf 5}, 24 (1967);
\bibitem{gut} M. Yoshimura, Phys. Rev. Lett. {\bf 41}, 281 (1978);
{\em ibid} {\bf 42}, 746 (E) (1979); Phys. Lett.{\bf B88}, 294 (1979);
S. Dimopoulos and L. Susskind, Phys. Rev. {\bf D18}, 4500 (1978);
D. Toussaint, S. B. treiman, F. Wilczek and A. Zee, Phys. Rev. {\bf D19},
1036 (1979); S. Weinberg, Phys. Rev. Lett. {\bf 42}, 850 (1979).
\bibitem{kuzmin} V. A. Kuzmin, V. A. Rubakov, and M. E. Shaposhnikov,
Phys. Lett. {\bf B155}, 36 (1985).
\bibitem{fukugita} M. Fukugita and T. Yanagida, Phys. Lett. {\bf B174}, 45 (1986).
\bibitem{luty} Markus A. Luty, Phys. Rev. {\bf D45}, 455 (1992);
M. Flanz, E. A. Paschos and U. Sarkar, Phys. Lett. {\bf B345}, 248 (1995);
L. Covi, E. Roulet and F. Vissani, Phys. Lett. {\bf B384}, 169 (1996);
W. Buchm\"{u}ller and M. Plumacher, Phys. Lett. {\bf B431}, 354 (1998).
For a review with an extensive list of references see 
W. Buchm\"{u}ller, R. D. Peccei, and T. Yanagida, Ann. Rev. Nucl. Part. Sci.
{\bf 55} 311 (2005) [arXiv:hep-ph/0502169].
\bibitem{lindner} K. Dick, M. Lindner, M. Ratz, and D. Wright, Phys. Rev. Lett. {\bf 84},
4039 (2000) [arXiv:hep-ph/99075062].
\bibitem{hung2} P. Q. Hung, Nucl. Phys. {\bf B747}, 55 (2006) [arXiv:hep-ph/0512282].
\bibitem{su2} P. Q. Hung, [arXiv:hep-ph/0504060].
\bibitem{snls} P. Astier {\em et al.}, Astron. Astrophys. {\bf 447} 31 (2006) 
[arXiv:astro-ph/0510447].
\bibitem{WMAP} D. N. Spergel {\em et al.}, ApJS, 170, 377 (2007) [arXiv:astro-ph/0603449].
\bibitem{hung4} Mehrdad Adibzadeh, P. Q. Hung, Nucl. Phys. {\bf B804}, 223 (2008) 
[arXiv:hep-ph/0805.3486v2].
\bibitem{hung3} P. Q. Hung, [arXiv:hep-ph/0604063].
\bibitem{kuzmin2} V. A. Kuzmin, Phys. Part. Nucl. {\bf 29} 257 (1998); 
Fiz.Elem.Chast.Atom.Yadra {\bf 29}, 637 (1998) [arXiv:hep-ph/9701269].
\bibitem{dodelson} Scott Dodelson and Lawrence M. Widrow, Phys. Rev. {\bf D42}, 326 (1990).
\bibitem{inflation} P.Q. Hung, Eduard Masso, Gabriel Zsembinszki, JCAP {\bf 0612}, 
004 (2006) [arXiv:astro-ph/0609777].
\bibitem{seesaw} R. N. Mohapatra and Goran Senjanovi\'{c}, Phys. Rev. {\bf D23}, 165 (1981).
\bibitem{susy} For a good summary of the definitions and properties of the two-component 
Weyl spinor formalism, see David Bailin and Alexander Love, 
{\em Supersymmetric Gauge Field Theory and String Theory},
Institute of Physics Publishing (2003)
\bibitem{majorana} K. M. Case, Phys. Rev. {\bf 107} 307 (1957).
\bibitem{wolfram} E. W. Kolb and S. Wolfram, Nucl. Phys. {\bf B172}, 224 (1980)
\bibitem{kolb} For a good pedagogical discussion, see E. W. Kolb and M. S. Turner, 
{\em The Early Universe}, Addison-Wesley Publishing Company (1990).
\bibitem{harvey} J. A. Harvey and M. S. Turner, Phys. Rev. {\bf D42} 3344 (1990)
\bibitem{majoranarule} A. Denner, H. Eck, O. Hahn and J. K\"{u}blbeck, Phys. Lett. {\bf B291}
278 (1992); Nucl. Phys. {\bf B387} 467 (1992)

\end{thebibliography}
\end{document}